\documentclass[]{IEEEtran}
\IEEEoverridecommandlockouts

\ifCLASSINFOpdf
\else
\fi

\usepackage{multirow}
\usepackage{cite}
\usepackage{stfloats}
\usepackage{color}
\usepackage{graphicx}
\usepackage{epstopdf}
\usepackage{slashbox}
\usepackage{float}
\usepackage{amssymb }
\usepackage[cmex10]{amsmath}
\usepackage{algorithm} 
\usepackage{algorithmic}
\usepackage{setspace}
\usepackage{amsfonts}
\usepackage{amsmath}
\usepackage{booktabs}
\usepackage{amsmath,bm}
\usepackage{tabularx}
\usepackage{makecell}
\usepackage{subfigure}

\def\BibTeX{{\rm B\kern-.05em{\sc i\kern-.025em b}\kern-.08em
    T\kern-.1667em\lower.7ex\hbox{E}\kern-.125emX}}
\usepackage{balance}

\begin{document}
\title{Symbol Detection in Multi-channel Multi-tag Ambient Backscatter
Communication Under IQ Imbalance}
\author{{Yuxin Li, Guangyue Lu, Yinghui Ye, Liqin Shi, and Daniel Benevides da Costa}
\thanks{Yuxin Li, Guangyue Lu, Yinghui Ye and Liqin Shi are with the Shaanxi Key Laboratory of
Information Communication Network and Security, Xi'an University of Posts
\& Telecommunications, Xi'an, Shaanxi 710121, China (e-mail: dreamyaurora@126.com, tonylugy@163.com, connectyyh@126.com, liqinshi@hotmail.com).}
\thanks{Daniel Benevides da Costa is with the Interdisciplinary Research Center for Communication Systems and Sensing (IRC-CSS), Department of Electrical Engineering, King Fahd University of Petroleum \& Minerals (KFUPM), Dhahran 31261, Saudi Arabia (email: danielbcosta@ieee.org).}
    \thanks{This work was supported in part by the National Natural Science Foundation of China under Grant 62471388 and 62301421, in part by the Innovation Capability Support Program of Shaanxi under Grant 2024ZC-KJXX-016, and in part by the Key Research and Development Program of Shaanxi under Grant 2024GX-ZDCYL-01-32.}
}
\maketitle

\begin{abstract}
Ambient backscatter communication (AmBC) offers low-cost and low-power connectivity for Internet of Things (IoT), where a backscatter tag (BT) modulates incident signals transmitted by an ambient radio frequency (RF) source and reflects them to its associated AmBC receiver. In multi-channel multi-tag AmBC, one of major challenges from the aspect of symbol detection is the image channel crosstalk, which is induced by the inevitable in-phase/quadrature (IQ) imbalance. To address this issue, in this paper, we study symbol detection in multi-channel multi-tag AmBC under IQ imbalance. Considering the differential encoding scheme at the BTs, we propose a novel symbol detection model that incorporates IQ imbalance parameters, the presence or absence of both the incident signal and the backscattered signal of the image channel. On this basis, considering an energy difference detector at the AmBC receiver, we derive the closed-form expressions for the bit error rate (BER) as well as the near-optimal detection threshold to minimize BER. However, calculating the near-optimal detection threshold requires prior information, such as the IQ imbalance parameters, the presence probability of the incident signal of the image channel and the backscattered signal of the image channel, the signal power of the ambient RF source, and the noise power, which are typically unknown to the AmBC receiver in practice. To eliminate the need for the prior information, we propose a threshold estimation method using the received samples. Numerical results indicate that under IQ imbalance, directly using the existing method leads to a significant degradation in BER performance. However, this degradation can be effectively mitigated by our derived detection threshold.

\end{abstract}


\IEEEpeerreviewmaketitle
\section{Introduction}
Internet of Things (IoT) for 6G green communication networks envisions a massive number of devices that are expected to be ultra-low power or near-zero power consumption and extremely low costs\cite{345,10130082,8368232,11005959,10764739}. To achieve this, ambient backscatter communication (AmBC) technology is  regarded as a promising solution, which empowers IoT devices as backscatter tags (BTs) to transmit information to the associated AmBC receiver over the signals emitted by the ambient radio frequency (RF) source. However, due to its spectrum-sharing behavior, backscattered signals suffer from the double path loss and interference from ambient RF signals. Thus, the recovery of the BT's information is challenging and it is crucial to design a reliable signal detector for the AmBC receiver to recover backscattered signals\cite{10353962}.

Considering that the backscattered signals are at a lower bit rate than the ambient RF signals, the authors in \cite{liu2013ambient} proposed an energy detector by leveraging the difference in communication rates. Then, the authors in \cite{lu2015signal} designed two detectors: one aimed at achieving the minimum bit error rate (BER), while the other was designed to ensure equal BER for detecting bit ``0'' and bit ``1''. However, obtaining the optimal detection thresholds in \cite{lu2015signal} required perfect channel state information (CSI) coefficients. Regarding this, a semi-coherent detector was proposed in \cite{8007328}, where the CSI coefficient can be estimated by jointly using a few pilot symbols. In addition, the differential encoding was employed at the BT to avoid channel estimation\cite{wang2016ambient,7769255}. Specifically, the authors in \cite{wang2016ambient} designed an energy difference  detector and derived the BER expression including its bounds at high signal-to-noise ratio (SNR). Following \cite{wang2016ambient}, the authors in \cite{7769255} designed a maximum likelihood (ML) detector based on the joint probability density function (PDF) of two consecutive differentially encoded symbols. Additionally, Manchester code and differential Manchester code were employed to encode information bits at the BT for both semi-coherent and non-coherent detection\cite{8329444}. Moreover, non-return-to-zero (NRZ) codes \cite{9242274}, space-time codes\cite{9430725}, and Miller codes\cite{9463672} were utilized at the BT to improve the performance of symbol detection.

The aforementioned studies \cite{liu2013ambient,lu2015signal,8007328,wang2016ambient,7769255,8329444,9242274,9430725,9463672}  have comprehensively investigated symbol detection in a single-channel and single-tag AmBC. In order to improve the total throughput of AmBC, multi-tag AmBC was studied, which enables BTs to backscatter the signals simultaneously. However, simultaneous transmission by multiple tags may cause a collision at the AmBC receiver and thus reduce the transmission efficiency. To mitigate such effects, parallel backscatter was proposed to recover BTs' information when there is a collision in \cite{angerer2010rfid}. However, it is based on the assumption that receivers can discriminate the sources of the two tags in the in-phase/quadrature (IQ) plane. To further distinguish multiple tag signals at the AmBC receiver, some works considered a time-division multiple
access (TDMA)-based scheme at BTs\cite{wang2018stackelberg,gu2022many}. In \cite{li2019price}, a multi-channel communication based on frequency-division multiple access (FDMA) was adopted in AmBC to enable simultaneous transmission of multiple BTs, where each BT was assigned a channel of ambient RF source. Although the authors in \cite{li2019price} introduced the multi-channel multi-tag AmBC network, the recovery of BTs' information at the AmBC receiver was not yet discussed. To recover the BTs' information in multi-channel multi-tag AmBC and to meet the low-cost nature of AmBC, it is necessary to employ a direct-conversion wideband multi-channel receiver for the multi-channel multi-tag AmBC.

It is noteworthy that the IQ imbalance associated with the RF front-end is inevitable\cite{7420754,4357467,1495892}, which can lead to the performance degradation of wireless communications. There were some efforts to address this issue in cognitive radio networks \cite{7463533} and full-duplex relay networks\cite{8085119}. However, research efforts aimed at resolving such a problem in AmBC are still at an early stage\cite{9415632,10005249}. In \cite{9415632}, the authors investigated the joint estimation of the channel, carrier frequency offset, and IQ imbalance at the AmBC receiver. In \cite{10005249}, the impacts of IQ imbalance on symbol detection for AmBC were discussed. However, the established symbol detection model and analysis in \cite{10005249} were limited to single-channel single-tag AmBC and considered only the IQ imbalance at the AmBC receiver. For a direct-conversion wideband multi-channel receiver, IQ imbalance will lead to image channel crosstalk, whereby each channel has image interference from its image channel. Meanwhile, the direct-conversion wideband multi-channel ambient RF source transmitter also suffers from image channel crosstalk caused by IQ imbalance, which has not yet received sufficient attention. Due to the image channel crosstalk, the received signal of the wideband AmBC receiver at a given channel consists of the RF source signals and backscattered signal of this channel, as well as the image interference signals of the image channel. As a result, the statistical distribution of the received signal under IQ imbalance will deviate from that of the received signal with the ideal transceiver. In this case, directly employing the detectors derived under ideal transceiver in a direct-conversion wideband multi-channel receiver may result in performance degradation. Thus, further investigation on the symbol detection for multi-channel multi-tag AmBC under IQ imbalance is required.

In this paper, we attempt to study the symbol detection in a multi-channel multi-tag AmBC network with non-ideal transceiver under IQ imbalance, where BTs are allowed to access the spectrum of RF source via FDMA. The main novelties and the contributions of this paper can be summarized as follows:

\begin{itemize}
  \item We propose a symbol detection framework for multi-channel multi-tag AmBC with non-ideal transceiver under IQ imbalance, where each BT backscatters its information in an FDMA manner. To represent the interference caused by IQ imbalance, we introduce two random variables that indicate the presence or absence of the incident signal and the backscattered signal of the image channel, respectively. To the best of our knowledge, this is the first time that such a multi-channel multi-tag AmBC symbol detection framework, with regard to IQ imbalance, is established in the open literature.
  \item We derive the closed-form BER expression as a function of the detection threshold, IQ imbalance parameters, the  presence probability of the incident signal of image channel and the backscattered signal of the image channel. The near-optimal detection threshold is also derived by minimizing BER. On this basis, the BER performance is analyzed, and the impacts of IQ imbalance on symbol detection for multi-channel multi-tag AmBC are investigated.
  \item Since the near-optimal detection threshold relies on parameters that are unavailable to the AmBC receiver, such as IQ imbalance, CSI coefficients, the presence probability of the incident signal of the image channel and the backscatter signal of the image channel, the signal power of the RF source, and noise power, we propose a practical method to estimate the threshold from the received samples for non-coherent symbol detection.
\end{itemize}

The remainder of this paper is organized as follows. In Section \uppercase\expandafter{\romannumeral2}, we introduce the system model for a multi-channel multi-tag AmBC network with non-ideal transceiver under IQ imbalance. In Section \uppercase\expandafter{\romannumeral3}, we derive the closed-form BER expression and near-optimal threshold, and we also design a practical method to estimate the near-optimal threshold. In Section \uppercase\expandafter{\romannumeral4}, the numerical results are presented. Finally, the conclusion is given in Section \uppercase\expandafter{\romannumeral5}.

\section{System Model}
\begin{figure}[t]
  \centering
  \includegraphics[width=0.45\textwidth]{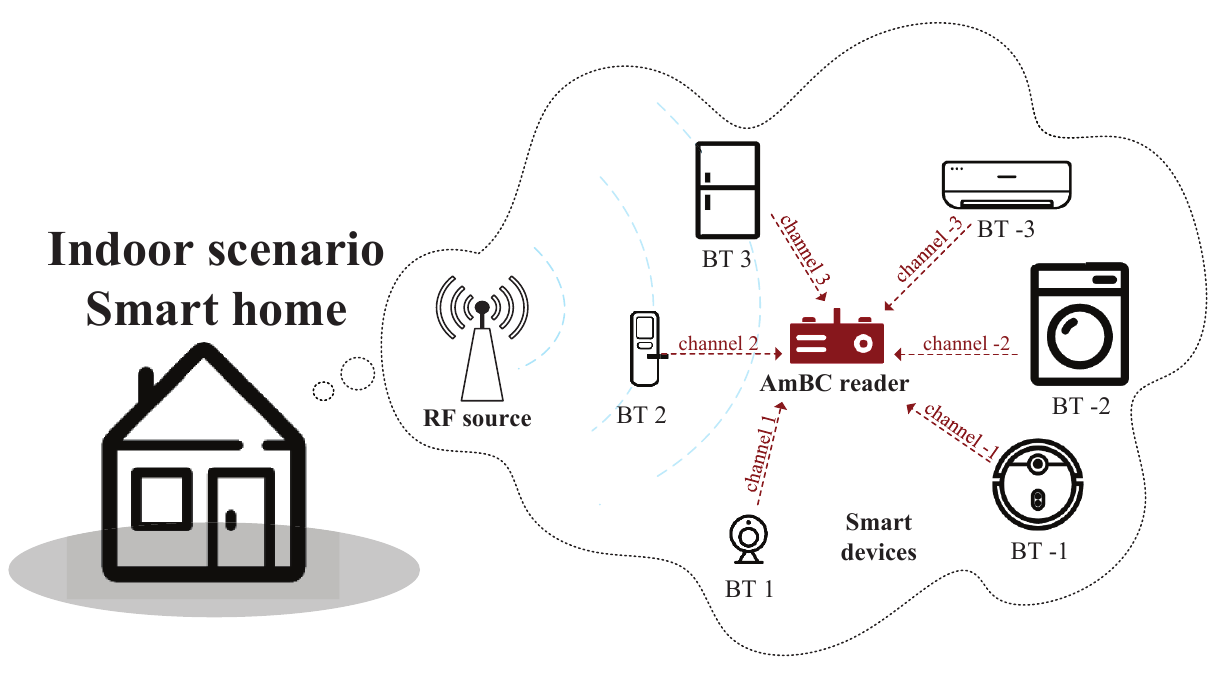}\\
  \caption{A typical indoor scenario of multi-channel multi-tag AmBC: Smart home.}\label{fig1}
  \vspace{-8pt}
\end{figure}

\begin{figure}[t]
  \centering
  \includegraphics[width=0.45\textwidth]{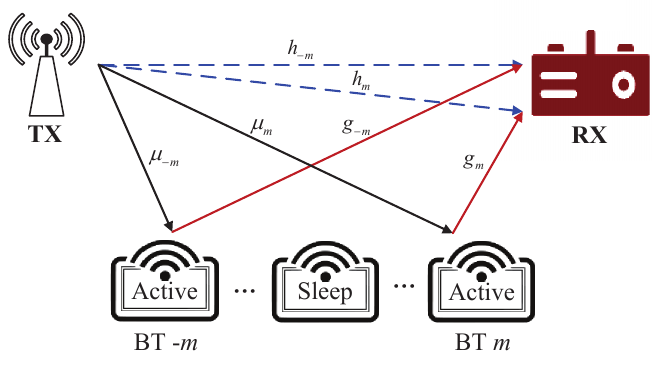}\\
  \caption{System model of multi-channel multi-tag AmBC. }\label{fig2}
  \vspace{-8pt}
\end{figure}

We consider a practical AmBC scenario involving a direct-conversion wideband ambient RF source transmitter (TX), $2M$ BTs, and a direct-conversion
wideband multi-channel AmBC receiver (RX), as depicted in Fig. \ref{fig1}. In this scenario, a Wi-Fi access point or a femto base station serves as the ambient RF source, and each smart device is attached with a BT to transmit data to the RX. The overall set of RF channels allocated to the TX is denoted by the enumeration $\left\{ m \mid -M \leq m \leq M, m \neq 0 \right\}$, which can be explicitly expressed as $S=\left\{ { -M, \dots, -1,1,\dots ,M} \right\}$. Each BT is allocated a dedicated channel from the set of $S$, ensuring a one-to-one correspondence between BTs and channels. Specifically, channel $m$ is assigned to BT $m$, and thus the set of BTs is given by $ B =\left\{ { -M, \dots, -1,1,\dots ,M} \right\}$. Fig. \ref{fig2} illustrates the system model, where $h_m$, $\mu_m$ and $g_m$ represent the complex channel coefficients between the TX and RX of channel $m$, between the TX and BT $m$ of channel $m$, and between BT $m$ and the RX of channel $m$, respectively. All channel coefficients are assumed to be flat block fading, remaining unchanged within each transmission block.

\begin{figure}[t]
\centering
\subfigure[Crosstalk at the TX]{
\label{fig3_a}
\includegraphics[width=0.22\textwidth]{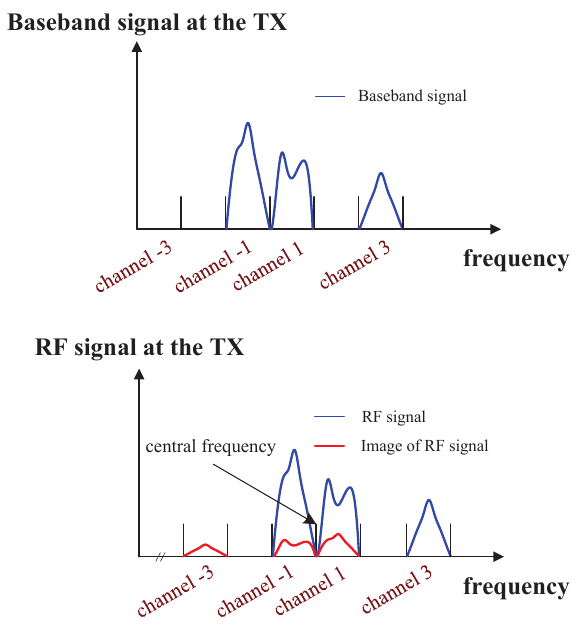}}
\subfigure[Crosstalk at the RX]{
\label{fig3_b}
\includegraphics[width=0.22\textwidth]{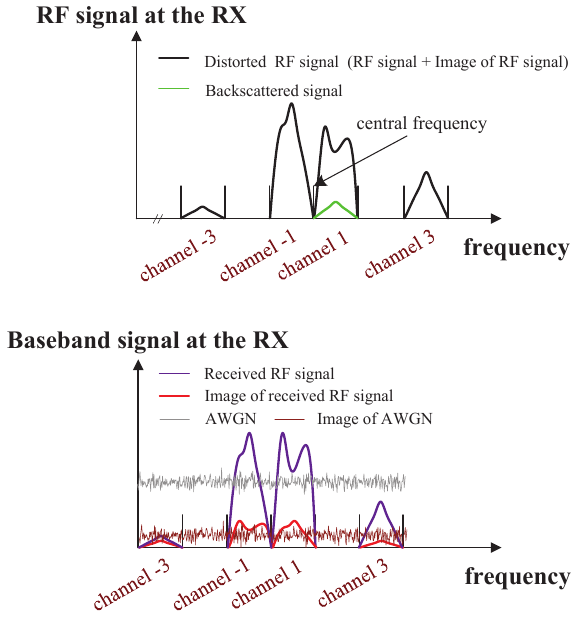}}
\caption{Illustration of the image channel crosstalk for
multi-channel multi-tag AmBC.}
\label{fig3}
\end{figure}

In this system, the signal of arbitrary channel $m$ may experience interference from the signal of the image channel $-m$ due to IQ imbalance at both the TX and RX. This phenomenon, referred to as image channel crosstalk\cite{4357467}, is illustrated in Fig. \ref{fig3}, with $M=3$ as an example. Fig. \ref{fig3_a} presents the baseband signal at the TX and demonstrates how IQ imbalance at the TX leads to image channel crosstalk. Specifically, the RF signal of channel $-1$ is affected by the image of the RF signal in channel $1$, and vice versa. Likewise, channel $-3$ experiences image interference from the RF signal of channel $3$. As a result, it is evident that the RF signal of channel $m$ can be distorted by the image interference in the presence of strong signal at its image channel. Subsequently, as shown in Fig. \ref{fig3_b}, an active BT $m$ modulates its information onto the distorted RF signal of channel $m$, which indicates that the backscattered signal is inherently influenced by IQ imbalance. For clarity, we assume that only BT $1$ is active, while all other BTs remain in sleep mode. Specifically, BT $1$ modulates its information onto the distorted RF signal of channel $1$. At the RX, the crosstalk behaves similarly to that at the TX. Consequently, the received baseband signal of channel $1$ consists of the distorted RF signal, the signal of BT 1 modulated onto the distorted RF signal and the image interference from the signal of channel $-1$. Hence, in the considered system, the BTs' information is modulated onto an already distorted RF signal, and the received RF signal at the RX is further impaired by the image interference, which pose a significant challenge to the recovery of BTs' information at the RX.

In this section, we propose a signal model for multi-channel multi-tag AmBC system under IQ imbalance. To this end, we first review the signal model under ideal transceivers as follows. The baseband received signal under ideal transceivers for $N$ consecutive samples of an arbitrary channel $m$ consists of the signals from the direct link, the backscatter link, and the noise, which can be given by
\begin{equation}
y_m^{{\rm{ideal}}}\left( n \right){\rm{ = }}\underbrace {{h_m}{p_m}\left( n \right)}_{{\rm{the\ direct\ link}}} + \underbrace {{\mu_m}{g_m}{B_m}\left( k \right){p_m}\left( n \right)}_{{\rm{the\ backscatter\ link}}} + \underbrace {{w_m}\left( n \right)}_{\rm{noise}},
\tag{1}
\label{eq1}
\end{equation}
where the superscript ``ideal'' denotes the assumption of ideal transceiver, $n=1,2,\dots,N$, ${p_{m}}(n)$ and $w_m(n)$ are the complex-valued baseband signal and the receiver noise of channel $m$ at time instant $n$, which are assumed to be complex Gaussian random variables, with distribution ${p_{m}}(n) \sim {\cal C}{\cal N}\left( {0,{P_s}} \right)$ and ${w_{m}}(n) \sim {\cal C}{\cal N}\left( {0,{\sigma_w^2}} \right)$, ${B_{m}}(k)$ is the $k_{th}$ transmission bit of BT $m$. The differential on-off keying (OOK) modulation is adopted at the BTs, where symbol ``0'' is encoded as two identical consecutive bits (``00'', ``11''), while symbol ``1'' is encoded as two alternating bits (``01'', ``10''). The output of the differential encoder for the $k_{th}$ symbol of BT $m$ is given by ${B_{m}}(k) = {A_{m}}(k) \otimes {B_{m}}(k - 1) $, where $A_m(k)$ denotes the pre-encoded symbol of BT $m$, and $\otimes$ represents addition modulo 2. Since the transmission rate of BTs is lower than that of TX, each BT symbol is assumed to remain unchanged over the $N$ consecutive TX samples.

In the case of IQ imbalance at the TX, the RF signals of arbitrary $m$ and its image channel $-m$ interfere with each other. According to \cite{schenk2008rf}, one typical source of IQ imbalance in the transceiver is modeled as phase and/or amplitude errors of local oscillator (LO) signals. Denote ${p_{-m}}(n)$ as the complex-valued baseband signal of channel $-m$ at time instant $n$, and let $\rho ^t$ and $\varphi ^t$ represent the amplitude and phase imbalance of the TX, respectively. Then, the distorted RF signal due to TX IQ imbalance for channel $m$ and its image channel $-m$ can be given by\cite{jalali2019cognitive}
\begin{equation}
s_m^{\rm{iq}}(n) = {\kappa _1^t{p_m}(n) + \kappa _2^t\xi p_{ - m}^*(n)} ,
\tag{2}
\label{eq2}
\end{equation}
\begin{equation}
s_{-m}^{\rm{iq}}(n) = {\kappa _1^t\xi {p_{-m}}(n) + \kappa _2^tp_{m}^*(n)} ,
\tag{3}
\label{eq3}
\end{equation}
where the superscript ``iq'' denotes IQ imbalance, $\ \xi = 1\ (\rm{or}\ 0 )$ indicates the presence (or absence) of the TX signal
of the channel $-m$, $*$ denotes the complex conjugate, and $\kappa _1^t$ and $\kappa _2^t$ are functions of $\rho ^r$ and $\varphi ^r$, which can be given by\cite{schenk2008rf}
\begin{equation}
\kappa _1^t = \frac{{1 + {\rho ^t}{e^{j{\varphi ^t}}}}}{2},
\tag{4}
\label{eq4}
\end{equation}
\begin{figure*}[!b]

\hrulefill
\begin{equation}
y_m^{{\rm{iq}}}(n) = \kappa _1^r\left( {{h_m}s_m^{{\rm{iq}}}\left( n \right) + g_m b_m^{{\rm{iq}}}\left( n \right) + {w_m}\left( n \right)} \right) + \kappa _2^r{\left( {{h_{ - m}}s_{ - m}^{{\rm{iq}}}\left( n \right) + g_{-m}b_{ - m}^{{\rm{iq}}}\left( n \right) + {w_{ - m}}\left( n \right)} \right)^*}.
\tag{8}
\label{eq8}
\end{equation}

\begin{equation}
y_m^{\rm{iq}}\left( n \right){\rm{  = }}\left\{ \begin{array}{l}
\kappa _1^r\left( {{h_m}\left( {\kappa _1^t{p_m}\left( n \right) + \kappa _2^t\xi {p_{ - m}}^*\left( n \right)} \right) + {w_m}\left( n \right)} \right)\\
 + \kappa _2^r{\left( {\left( {{h_{ - m}} + \eta {\mu _{ - m}}{g_{ - m}}} \right)\left( {\kappa _1^t\xi {p_{ - m}}\left( n \right) + \kappa _2^t{p_m}^*\left( n \right)} \right) + {w_{ - m}}\left( n \right)} \right)^ * },\ {B_m}(k) = 0\\
\kappa _1^r\left( {\left( {{h_m} + {\mu _m}{g_m}} \right)\left( {\kappa _1^t{p_m}\left( n \right) + \kappa _2^t\xi {p_{ - m}}^*\left( n \right)} \right) + {w_m}\left( n \right)} \right)\\
+ \kappa _2^r{\left( {\left( {{h_{ - m}} + \eta {\mu _{ - m}}{g_{ - m}}} \right)\left( {\kappa _1^t\xi {p_{ - m}}\left( n \right) + \kappa _2^t{p_m}^*\left( n \right)} \right) + {w_{ - m}}\left( n \right)} \right)^ * },\ {B_m}(k) = 1
\end{array} \right.\tag{11}
\label{eq11}
\end{equation}
\begin{equation}
\left\{ {{\begin{array}{*{20}{l}}
\mathbb{E}{\left( {\Gamma _{m,k\left| {\xi ,\eta } \right.}^{{\rm{iq}}}\left| {B_m(k) = 0} \right.} \right) = {{\left| {\kappa _1^r\kappa _1^t{h_m} + \kappa _2^r{{\left( {\kappa _2^t} \right)}^*}{{\left( {{h_{ - m}} + \eta {\mu _{ - m}}{g_{ - m}}} \right)}^*}} \right|}^2}{P_s}}\\
{\qquad \qquad \qquad \qquad \qquad  \quad + {{\left| {\kappa _1^r\kappa _2^t{h_m} + \kappa _2^r{{\left( {\kappa _1^t} \right)}^*}{{\left( {{h_{ - m}} + \eta {\mu _{ - m}}{g_{ - m}}} \right)}^*}} \right|}^2}\xi {P_s} + \left( {{{\left| {\kappa _1^r} \right|}^2} + {{\left| {\kappa _2^r} \right|}^2}} \right)\sigma _w^2}\\
\mathbb{E}{\left( {\Gamma _{m,k\left| {\xi ,\eta } \right.}^{{\rm{iq}}}\left| {B_m(k) = 1} \right.} \right) = {{\left| {\kappa _1^r\kappa _1^t\left( {{\mu_m} + {\mu _m}{g_m}} \right) + \kappa _2^r{{\left( {\kappa _2^t} \right)}^*}{{\left( {{h_{ - m}} + \eta {\mu _{ - m}}{g_{ - m}}} \right)}^*}} \right|}^2}{P_s}}\\
{\qquad \qquad \qquad \qquad \qquad  \quad + {{\left| {\kappa _1^r\kappa _2^t\left( {{h_m} + {\mu _m}{g_m}} \right) + \kappa _2^r{{\left( {\kappa _1^t} \right)}^*}{{\left( {{h_{ - m}} + \eta {\mu _{ - m}}{g_{ - m}}} \right)}^*}} \right|}^2}\xi {P_s} + \left( {{{\left| {\kappa _1^r} \right|}^2} + {{\left| {\kappa _2^r} \right|}^2}} \right)\sigma _w^2}
\end{array}}} \right.
\tag{14}
\label{eq14}
\end{equation}

\begin{equation}
\left\{ \begin{array}{l}
\mathbb{D}\left( {\Gamma _{m,k\left| {\xi ,\eta } \right.}^{{\rm{iq}}}\left| {B_m(k) = 0} \right.} \right) = \frac{{{{\left| {\mathbb{E}\left( {\Gamma _{m,k\left| {\xi ,\eta } \right.}^{{\rm{iq}}}\left| B_m(k) = 0 \right.} \right)} \right|}^2}}}{N}\\
\mathbb{D}\left( {\Gamma _{m,k\left| {\xi ,\eta } \right.}^{{\rm{iq}}}\left| {B_m(k) = 1} \right.} \right) = \frac{{{{\left| {\mathbb{E}\left( {\Gamma _{m,k\left| {\xi ,\eta } \right.}^{{\rm{iq}}}\left| B_m(k) = 1 \right.} \right)} \right|}^2}}}{N}
\end{array} \right.
\tag{15}
\label{eq15}
\end{equation}

\end{figure*}
\begin{equation}
{\rm{ }}\kappa _2^t = \frac{{1 - {\rho ^t}{e^{j{\varphi ^t}}}}}{2}.
\tag{5}
\label{eq5}
\end{equation}

Then, the backscattered signal of BT $m$ modulated over the distorted RF signal of channel $m$ can be written as\footnote{The BT does not include RF components \cite{9866050}, thereby avoiding issues related to IQ imbalance.}
\begin{equation}
b_m^{\rm{iq}}(n) = {\mu _m}s_m^{\rm{iq}}(n){B_m}(k).
\tag{6}
\label{eq6}
\end{equation}

Let $\ \eta = 1\ (\rm{or}\ 0 )$ indicate the presence (or absence) of the backscattered signal of BT $-m$, and ${B_{-m}}(k)$ be the $k_{th}$ symbol of the BT $-m$. The backscattered signal of BT $-m$ modulated over the distorted RF signal of channel $-m$ can be denoted as
\begin{equation}
\begin{array}{l}
b_{ - m}^{\rm{iq}}(n) = \eta{\mu _{ - m}}s_{-m}^{\rm{iq}}(n){B_{ - m}}(k).
\end{array}
\tag{7}
\label{eq7}
\end{equation}

Considering IQ imbalance at the RX, the received baseband samples $y_{m}^{\rm{iq}}(n)$ can be mathematically expressed as (\ref{eq8}). Here, $w_{-m}(n)$ is the receiver noise of channel $-m$ at time instant $n$. With $\rho ^r$ and $\varphi ^r$ representing the amplitude and phase imbalances at the RX, the coefficients $\kappa _1^r$ and $\kappa _2^r$ are\cite{schenk2008rf}
\begin{equation}
\kappa _1^r = \frac{{1 + {\rho ^r}{e^{-j{\varphi ^r}}}}}{2},
\tag{9}
\label{eq9}
\end{equation}
\begin{equation}
{\rm{ }}\kappa _2^r = \frac{{1 - {\rho ^r}{e^{j{\varphi ^r}}}}}{2}.
\tag{10}
\label{eq10}
\end{equation}
Then, by substituting (\ref{eq2}), (\ref{eq3}), (\ref{eq6}) and (\ref{eq7}) into (\ref{eq8}), the $n_{th}$ received signal of channel $m$ under IQ imbalance can be rewritten as (\ref{eq11}).

{\it{Remark 1:}} When $\rho^t=1$, $\rho ^r=1$, $\varphi^t=0$ and $\varphi ^r=0$, the proposed signal model in (\ref{eq8}) under IQ imbalance degrades to the ideal signal model without IQ imbalance, as given in (\ref{eq1}).

{\it{Remark 2:}} Due to image channel crosstalk, the statistical distribution of the received baseband signal under IQ imbalance, $y_m^{\rm{iq}}$, deviates from that of the received baseband signal with the ideal transceiver, $y_m^{\rm{ideal}}$.

\section{Symbol Detection under IQ Imbalance}

\subsection{BER and Near-optimal Detection Threshold}
Under IQ imbalance, the test statistic based on energy difference detector is given by
\begin{equation}
\left| T_{m,k}^{\rm{iq}} \right| =  \left|{\Gamma _{m,k}^{\rm{iq}} - \Gamma _{m,k-1}^{\rm{iq}}}\right|,
\tag{12}
\label{eq12}
\end{equation}
where ${\Gamma _{m,k}^{{\rm{iq}}}} = \frac{1}{N}\sum\nolimits_{n = (k - 1)N + 1}^{kN} {{{\left| {y_m^{\rm{iq}}(n)} \right|}^2}}$, is the average power of the $N$ consecutive samples for the $k_{th}$ bit of channel $m$. Then the detector for multi-channel
multi-tag AmBC under IQ imbalance can be expressed as,
\begin{equation}
\left\{ \begin{array}{l}
{{\hat A}_m}(k) = 1,\ \left| T_{m,k}^{\rm{iq}}\right| \ge \gamma _{m,k}^{\rm{iq}}\\[6pt]
{{\hat A}_m}(k) = 0,\ \left| T_{m,k}^{\rm{iq}}\right| <  \gamma _{m,k}^{\rm{iq}}
\end{array} \right.,
\tag{13}
\label{eq13}
\end{equation}
where ${{\hat A}_m}(k)$ is the estimated value of ${{ A}_m}(k)$, and $\gamma _{m,k}^{\rm{iq}}$ is the detection threshold under IQ imbalance.

As reported in \cite{wang2016ambient}, the near-optimal detection threshold and BER for the energy difference detector exhibit a respective mapping relationship with the mean and variance of the average power of the received signal. To obtain the required mean and variance of ${\Gamma _{m,k}^{{\rm{iq}}}}$, it is necessary to derive the distribution of ${\Gamma _{m,k\left| {\xi,\eta} \right.}^{{\rm{iq}}}}$ conditioned on the presence or absence of the TX signal of the channel $- m$, $\xi$, and the presence or absence of the backscattered signal of the BT $-m$, $\eta$. Towards this end, Theorem 1 is provided.

\emph{\textit{Theorem 1:}} The distribution of ${\Gamma _{m,k\left| {\xi,\eta} \right.}^{{\rm{iq}}}}$ conditioned on $\xi$ and $\eta$, can be approximated by a Gaussian distribution with mean and variance given in (\ref{eq14}) and (\ref{eq15}), respectively, for large $N$.

\emph{Proof:} The detailed proof is presented in Appendix A.\hfill {$\blacksquare $}

From (\ref{eq14}) and (\ref{eq15}), we can see that the mean and variance of ${\Gamma _{m,k\left| {\xi,\eta} \right.}^{{\rm{iq}}}}$ are both related to the parameters $\xi$ and $\eta$. To obtain a general performance analysis, we treat $\xi$ and $\eta$ as random variables, which allows us to explore the conditional distribution under the joint distribution of $\xi$ and $\eta$. To this end, we assume that $\xi$, the parameter of TX signal at image channel $-m$, follows Bernoulli distribution\footnote{It is widely acknowledged that the arrivals of signals at channel are Bernoulli random variables\cite{5740629}.} with success parameter $0 \le q \le 1$, i.e., $\Pr \left( {\xi  = 1} \right) = q$ and $\Pr \left( {\xi  = 0} \right) = 1-q$. To further study the parameter $\eta$, we assume that BT is in sleep mode to save energy when there is no data to be transmitted, as in \cite{8103807}. Once there are enough data to be transmitted, BT will enter active state with parameter $0 \le v \le 1$, i.e., $\Pr \left( {\rm{state  = active}} \right) = v$ and $\Pr \left( \rm{{state  = sleep}} \right) = 1-v$ for any BT\footnote {This can be acquired through a series of trials and experiments in practical scenarios.}, which can be considered as Bernoulli distribution with success parameter $0 \le v \le 1$. On this basis, the distribution of $\eta$ is provided in Lemma 2.

\emph{\textit{Lemma 2:}} The presence or absence of backscattered signal of BT $-m$, denoted by $\eta$, follows a Bernoulli distribution with a success parameter $0 \le \frac{v}{2} \le \frac{1}{2}$, i.e., $\Pr \left( {\eta  = 1} \right) = \frac{v}{2}$ and $\Pr \left( {\eta  = 0} \right) = 1-\frac{v}{2}$.

\emph{Proof:} The detailed proof is presented in Appendix B.\hfill {$\blacksquare $}

{\it{Remark 2:}} Evidently, the random variables $\xi$ and $\eta$ are mutually independent.
\begin{figure*}[!b]
\hrulefill
\begin{equation}
\left\{ \begin{array}{l}
\mathbb{E}\left( {\Gamma _{m,k}^{{\rm{iq}}}\left| {B_m(k) = 0} \right.} \right) = \mathbb{E}\left( {\Gamma _{m,k\left| {\xi  = 0,\eta  = 0} \right.}^{{\rm{iq}}}\left| {B_m(k) = 0} \right.} \right){\Pr \left( {\xi  = 0,\eta  = 0} \right)} + \mathbb{E}\left( {\Gamma _{m,k\left| {\xi  = 0,\eta  = 1} \right.}^{{\rm{iq}}}\left| {B_m(k) = 0} \right.} \right){\Pr \left( {\xi  = 0,\eta  = 1} \right)}\\
 \qquad \qquad \qquad \qquad \qquad + \mathbb{E}\left( {\Gamma _{m,k\left| {\xi  = 1,\eta  = 0} \right.}^{{\rm{iq}}}\left| {B_m(k) = 0} \right.} \right){\Pr \left( {\xi  = 1,\eta  = 0} \right)} + \mathbb{E}\left( {\Gamma _{m,k\left| {\xi  = 1,\eta  = 1} \right.}^{{\rm{iq}}}\left| {B_m(k) = 0} \right.} \right){\Pr \left( {\xi  = 1,\eta  = 1} \right)}\\
\mathbb{E}\left( {\Gamma _{m,k}^{{\rm{iq}}}\left| {B_m(k) = 1} \right.} \right) = \mathbb{E}\left( {\Gamma _{m,k\left| {\xi  = 0,\eta  = 0} \right.}^{{\rm{iq}}}\left| {B_m(k) = 1} \right.} \right){\Pr \left( {\xi  = 0,\eta  = 0} \right)} + \mathbb{E}\left( {\Gamma _{m,k\left| {\xi  = 0,\eta  = 1} \right.}^{{\rm{iq}}}\left| {B_m(k) = 1} \right.} \right){\Pr \left( {\xi  = 0,\eta  = 1} \right)}\\
 \qquad \qquad \qquad \qquad \qquad + \mathbb{E}\left( {\Gamma _{m,k\left| {\xi  = 1,\eta  = 0} \right.}^{{\rm{iq}}}\left| {B_m(k) = 1} \right.} \right){\Pr \left( {\xi  = 1,\eta  = 0} \right)} + \mathbb{E}\left( {\Gamma _{m,k\left| {\xi  = 1,\eta  = 1} \right.}^{{\rm{iq}}}\left| {B_m(k) = 1} \right.} \right){\Pr \left( {\xi  = 1,\eta  = 1} \right)}
\end{array} \right.
\tag{24}
\label{eq24}
\end{equation}
\begin{equation}
T_{m,k}^{\rm{iq}} \sim \left\{ \begin{array}{l}
\left. \begin{array}{l}
{\cal N}\left( {0,2\Delta _0^2} \right),{B_m}(k) = 0,{B_m}(k - 1) = 0\\
{\cal N}\left( {0,2\Delta _1^2} \right),{B_m}(k) = 1,{B_m}(k - 1) = 1
\end{array} \right\}{A_m}(k) = 0\\
\left. \begin{array}{l}
{\cal N}\left( {\vartheta _{m,k}^{{\rm{iq}}},\Delta _ + ^2} \right),{B_m}(k) = 1,{B_m}(k - 1) = 0\\
{\cal N}\left( { - \vartheta _{m,k}^{{\rm{iq}}},\Delta _ + ^2} \right),{B_m}(k) = 0,{B_m}(k - 1) = 1
\end{array} \right\}{A_m}(k) = 1
\end{array} \right.
\tag{26}
\label{eq26}
\end{equation}
\end{figure*}

{\it{Remark 3:}} Given that $v=1$ indicates BT is in active mode and the probability of bit ``1'' and bit ``0'' are equal to $\frac{1}{2}$, the presence probability of backscattered signal of BT $-m$ is no more than $\frac{1}{2}$.

To conduct a general analysis of the distribution of ${\Gamma _{m,k}^{{\rm{iq}}}}$, we begin by analyzing the mean of ${\Gamma _{m,k\left| {\xi,\eta} \right.}^{{\rm{iq}}}}$ for specific values of $\xi$ and $\eta$ followed by averaging over all possible cases. Remark 2 points out that $\xi$ and $\eta$ are mutually independent, yielding four possible cases that are $\xi  = 0$ and $\eta  = 0$, $\xi  = 0$ and $\eta  = 1$, $\xi  = 1$ and $\eta  = 0$, $\xi  = $ and $\eta  = 1$ by combining two random variables arbitrarily. In the following, we analyze the probability and mean of ${\Gamma _{m,k\left| {\xi,\eta} \right.}^{{\rm{iq}}}}$  for each case.

\emph{\textit{Case 1:}} $\xi  = 0$ and $\eta  = 0$

The probability for case 1 is

\begin{align}\label{eq16}
\Pr \left( {\xi  = 0,\eta  = 0} \right) &= \Pr \left( {\xi  = 0} \right)\Pr \left( {\eta  = 0} \right)\notag\\
 &= \left( {1 - q} \right)\left( {1 - \frac{v}{2}} \right).\tag{16}
\end{align}

The mean of ${\Gamma _{m,k\left| {\xi  = 0,\eta  = 0} \right.}^{{\rm{iq}}}}$ is given by
\begin{equation}
\left\{ \begin{array}{l}
\quad\mathbb{E}\left( {\Gamma _{m,k\left| {\xi  = 0,\eta  = 0} \right.}^{{\rm{iq}}}\left| {B_m(k) = 0} \right.} \right)\\
 = {\left| {\kappa _1^r\kappa _1^t{h_m} + \kappa _2^r{{\left( {\kappa _2^t} \right)}^*}{{\left( {{h_{ - m}}} \right)}^*}} \right|^2}{P_s}\\
\quad + \left( {{{\left| {\kappa _1^r} \right|}^2} + {{\left| {\kappa _2^r} \right|}^2}} \right)\sigma _w^2\\
\quad\mathbb{E}\left( {\Gamma _{m,k\left| {\xi  = 0,\eta  = 0} \right.}^{{\rm{iq}}}\left| {B_m(k) = 1} \right.} \right)\\
 = {\left| {\kappa _1^r\kappa _1^t\left( {{h_m} + {\mu _m}{g_m}} \right) + \kappa _2^r{{\left( {\kappa _2^t} \right)}^*}{{\left( {{\mu_{ - m}}} \right)}^*}} \right|^2}{P_s}\\
\quad + \left( {{{\left| {\kappa _1^r} \right|}^2} + {{\left| {\kappa _2^r} \right|}^2}} \right)\sigma _w^2
\end{array} \right.
\tag{17}
\label{eq17}
\end{equation}

\emph{\textit{Case 2:}} $\xi  = 0$ and $\eta  = 1$

The probability for case 2 is
\begin{align}\label{eq18}
\Pr \left( {\xi  = 0,\eta  = 1} \right) &= \Pr \left( {\xi  = 0} \right)\Pr \left( {\eta  = 1} \right)\notag\\
 &=\frac{{\left( {1 - q} \right)v}}{2}.\tag{18}
\end{align}

The mean of ${\Gamma _{m,k\left| {\xi  = 0,\eta  = 1} \right.}^{{\rm{iq}}}}$ reads
\begin{equation}
\left\{ \begin{array}{l}
\quad \mathbb{E}\left( {\Gamma _{m,k\left| {\xi  = 0,\eta  = 1} \right.}^{{\rm{iq}}}\left| {B_m(k) = 0} \right.} \right)\\
 = {\left| {\kappa _1^r\kappa _1^t{h_m} + \kappa _2^r{{\left( {\kappa _2^t} \right)}^*}{{\left( {{h_{ - m}} + {\mu _{ - m}}{g_{ - m}}} \right)}^*}} \right|^2}{P_s}\\
\quad + \left( {{{\left| {\kappa _1^r} \right|}^2} + {{\left| {\kappa _2^r} \right|}^2}} \right)\sigma _w^2\\
\quad \mathbb{E}\left( {\Gamma _{m,k\left| {\xi  = 0,\eta  = 1} \right.}^{{\rm{iq}}}\left| {B_m(k) = 1} \right.} \right)\\
={\left| {\kappa _1^r\kappa _1^t\left( {{h_m} + {\mu _m}{g_m}} \right) + \kappa _2^r{{\left( {\kappa _2^t} \right)}^*}{{\left( {{h_{ - m}} + {\mu _{ - m}}{g_{ - m}}} \right)}^*}} \right|^2}{P_s}\\
\quad + \left( {{{\left| {\kappa _1^r} \right|}^2} + {{\left| {\kappa _2^r} \right|}^2}} \right)\sigma _w^2
\end{array} \right.
\tag{19}
\label{eq19}
\end{equation}

\emph{\textit{Case 3:}} $\xi  = 1$ and $\eta  = 0$

The probability for case 3 is
\begin{align}\label{eq20}
\Pr \left( {\xi  = 1,\eta  = 0} \right) &= \Pr \left( {\xi  = 1} \right)\Pr \left( {\eta  = 0} \right)\notag\\
 &=q\left( {1 - \frac{v}{2}} \right).\tag{20}
\end{align}

The mean of ${\Gamma _{m,k\left| {\xi  = 1,\eta  = 0} \right.}^{{\rm{iq}}}}$ is denoted by
\begin{equation}
\left\{ \begin{array}{l}
\quad\mathbb{E}\left( {\Gamma _{m,k\left| {\xi  = 1,\eta  = 0} \right.}^{{\rm{iq}}}\left| {B_m(k) = 0} \right.} \right) \\
={\left| {\kappa _1^r\kappa _1^t{h_m} + \kappa _2^r{{\left( {\kappa _2^t} \right)}^*}{{\left( {{h_{ - m}}} \right)}^*}} \right|^2}{P_s}\\
\quad + {\left| {\kappa _1^r\kappa _2^t{h_m} + \kappa _2^r{{\left( {\kappa _1^t} \right)}^*}{{\left( {{h_{ - m}}} \right)}^*}} \right|^2}{P_s}\\
\quad+ \left( {{{\left| {\kappa _1^r} \right|}^2} + {{\left| {\kappa _2^r} \right|}^2}} \right)\sigma _w^2\\
\quad\mathbb{E}\left( {\Gamma _{m,k\left| {\xi  = 1,\eta  = 0} \right.}^{{\rm{iq}}}\left| {B_m(k) = 1} \right.} \right)\\
 = {\left| {\kappa _1^r\kappa _1^t\left( {{h_m} + {\mu _m}{g_m}} \right) + \kappa _2^r{{\left( {\kappa _2^t} \right)}^*}{{\left( {{h_{ - m}}} \right)}^*}} \right|^2}{P_s}\\
\quad + {\left| {\kappa _1^r\kappa _2^t\left( {{h_m} + {\mu _m}{g_m}} \right) + \kappa _2^r{{\left( {\kappa _1^t} \right)}^*}{{\left( {{h_{ - m}}} \right)}^*}} \right|^2}{P_s}\\
\quad+ \left( {{{\left| {\kappa _1^r} \right|}^2} + {{\left| {\kappa _2^r} \right|}^2}} \right)\sigma _w^2
\end{array} \right.
\tag{21}
\label{eq21}
\end{equation}

\emph{\textit{Case 4:}} $\xi  = 1$ and $\eta  = 1$

The probability for case 4 is
\begin{align}\label{eq22}
\Pr \left( {\xi  = 1,\eta  = 1} \right) &= \Pr \left( {\xi  = 1} \right)\Pr \left( {\eta  = 1} \right)\notag\\
 &=\frac{{qv}}{2}.\tag{22}
\end{align}

The mean of ${\Gamma _{m,k\left| {\xi  = 1,\eta  = 1} \right.}^{{\rm{iq}}}}$ can be
\begin{equation}
\left\{ \begin{array}{l}
\quad\mathbb{E}\left( {\Gamma _{m,k\left| {\xi  = 1,\eta  = 1} \right.}^{{\rm{iq}}}\left| {B_m(k) = 0} \right.} \right)\\
 = {\left| {\kappa _1^r\kappa _1^t\left( {{h_m} + {\mu _m}{g_m}} \right) + \kappa _2^r{{\left( {\kappa _2^t} \right)}^*}{{\left( {{h_{ - m}} + {\mu _{ - m}}{g_{ - m}}} \right)}^*}} \right|^2}{P_s}\\
 \quad+ {\left| {\kappa _1^r\kappa _2^t\left( {{h_m} + {\mu _m}{g_m}} \right) + \kappa _2^r{{\left( {\kappa _1^t} \right)}^*}{{\left( {{h_{ - m}} + {\mu _{ - m}}{g_{ - m}}} \right)}^*}} \right|^2}{P_s}\\
 \quad+ \left( {{{\left| {\kappa _1^r} \right|}^2} + {{\left| {\kappa _2^r} \right|}^2}} \right)\sigma _w^2\\
\quad\mathbb{E}\left( {\Gamma _{m,k\left| {\xi  = 1,\eta  = 1} \right.}^{{\rm{iq}}}\left| {B_m(k) = 1} \right.} \right)\\
 = {\left| {\kappa _1^r\kappa _1^t{h_m} + \kappa _2^r{{\left( {\kappa _2^t} \right)}^*}{{\left( {{h_{ - m}} + {\mu _{ - m}}{g_{ - m}}} \right)}^*}} \right|^2}{P_s}\\
 \quad+ {\left| {\kappa _1^r\kappa _2^t{h_m} + \kappa _2^r{{\left( {\kappa _1^t} \right)}^*}{{\left( {{h_{ - m}} + {\mu _{ - m}}{g_{ - m}}} \right)}^*}} \right|^2}{P_s}\\
\quad + \left( {{{\left| {\kappa _1^r} \right|}^2} + {{\left| {\kappa _2^r} \right|}^2}} \right)\sigma _w^2
\end{array} \right.
\label{eq23}
\tag{23}
\end{equation}

For the general case, mean of $\Gamma _{m,k}^{{\rm{iq}}}$ is computed by taking the expectation over the four cases, where the mean values from (\ref{eq17}), (\ref{eq19}), (\ref{eq21}), and (\ref{eq23}) are weighted by their corresponding probabilities, as shown in (\ref{eq24}). Furthermore, the variance of $\Gamma _{m,k}^{{\rm{iq}}}$ can be approximated as,

\begin{equation}
\left\{ \begin{array}{l}
\mathbb{D}\left( {\Gamma _{m,k}^{{\rm{iq}}}\left| {B_m(k) = 0} \right.} \right) = \frac{{\mathbb{E}^2{{ {\left( {\Gamma _{m,k}^{{\rm{iq}}}\left| {B_m(k) = 0} \right.} \right)} }}}}{N}\\
\mathbb{D}\left( {\Gamma _{m,k}^{{\rm{iq}}}\left| {B_m(k) = 1} \right.} \right) = \frac{{\mathbb{E}^2{{ {\left( {\Gamma _{m,k}^{{\rm{iq}}}\left| {B_m(k) = 1} \right.} \right)} }}}}{N}
\end{array} \right.
\tag{25}
\label{eq25}
\end{equation}

Given that $\Gamma _{m,k}^{{\rm{iq}}}$ follows a Gaussian distribution with a known mean and variance, and applying the closure property of Gaussian distribution\cite{papoulis2002probability}, ${T_{m,k}^{\rm{iq}}}$ can be approximated as Gaussian distribution, as formulated in (\ref{eq26}). The parameters, $\vartheta _{m,k}^{\rm{iq}}$, $\Delta _0^2$, and $\Delta _1^2$ , can be respectively given by,
\begin{equation}
\vartheta _{m,k}^{{\rm{iq}}} = \mathbb{E}\left( {\Gamma _{m,k}^{{\rm{iq}}}\left| {B_m(k) = 1} \right.} \right) - \mathbb{E}\left( {\Gamma _{m,k}^{{\rm{iq}}}\left| {B_m(k) = 0} \right.} \right),
\tag{27}
\label{eq27}
\end{equation}
\begin{equation}
\Delta _0^2 = \mathbb{D}\left( {\Gamma _{m,k}^{{\rm{iq}}}\left| {B_m(k) = 0} \right.} \right),
\tag{28}
\label{eq28}
\end{equation}
\begin{equation}
\Delta _1^2 = \mathbb{D}\left( {\Gamma _{m,k}^{{\rm{iq}}}\left| {B_m(k) = 1} \right.} \right).
\tag{29}
\label{eq29}
\end{equation}

Let $f_0(x)$ and $f_1(x)$ represent the PDFs of $T _{m,k}^{{\rm{iq}}}$ for $A_m(k)=0$ and $A_m(k)=1$, respectively, thus we have
\begin{align}
{f_0}(x) &= \frac{1}{2} \left(
\frac{1}{{\sqrt{4\pi \Delta_0^2}}} \exp \left( -\frac{x^2}{4\Delta_0^2} \right) \right. \notag \\
&\quad + \left. \frac{1}{{\sqrt{4\pi \Delta_1^2}}} \exp \left( -\frac{x^2}{4\Delta_1^2} \right)
\right),
\tag{30}
\label{eq30}
\end{align}
\begin{align}
{f_1}(x) &= \frac{1}{2} \left(
\frac{1}{{\sqrt{2\pi \Delta_+^2}}} \exp \left( -\frac{{{{\left( {x - \vartheta _{m,k}^{{\rm{iq}}}} \right)}^2}}}{2 \Delta_+^2} \right) \right. \notag \\
&\quad + \left. \frac{1}{{\sqrt{2\pi \Delta_+^2}}} \exp \left( -\frac{{{{\left( {x + \vartheta _{m,k}^{{\rm{iq}}}} \right)}^2}}}{2\Delta_+^2} \right)
\right).
\tag{31}
\label{eq31}
\end{align}

Then, the optimal threshold based on the ML criterion can be obtained by solving $f_0(x) = f_1(x)$, which ensures a minimum BER. However, obtaining a closed-form solution is intractable. Therefore, we resort to an approximation to derive a near-optimal threshold. Following \cite{wang2016ambient}, we approximate $f_0(x)$ as $\tilde{f}_0(x) = \frac{1}{\sqrt{2\pi \Delta_+^2}} \exp\!\left(-\frac{x^2}{2\Delta_+^2}\right)$,
based on which the near-optimal detection threshold for the RX under IQ imbalance can be given by,
\begin{align}
\gamma _{m,k}^{{\rm{iq}}} =& \frac{{\left| {\vartheta _{m,k}^{{\rm{iq}}}} \right|}}{2} +\frac{{\Delta _ + ^2}}{{\left| {\vartheta _{m,k}^{{\rm{iq}}}} \right|}}\notag \\
 &\times\ln \left( {1 + \sqrt {1 - \exp \left( { - \frac{{{{\left( {\vartheta _{m,k}^{{\rm{iq}}}} \right)}^2}}}{{\Delta _ + ^2}}} \right)} } \right),
\tag{32}
\label{eq32}
\end{align}
where $\Delta_+^2 = \Delta_0^2 + \Delta_1^2$. And the BER can be expressed as
\begin{align}\label{eq33}
\mathbb{P}_{{\rm{BER}}}^{{\rm{iq}}} = &\Pr \left( {A_m(k) = 1} \right)\Pr \left( {\hat A_m(k) = 0\left| {A_m(k) = 1} \right.} \right)\notag\\
\quad &+ \Pr \left( {A_m(k) = 0} \right)\Pr \left( {\hat A_m(k) = 1\left| {A_m(k) = 0} \right.} \right)\notag\\
= &\frac{1}{2}Q\left( {\frac{{\gamma _{m,k}^{\rm{iq}}}}{{\sqrt {2\Delta _0^2} }}} \right) + \frac{1}{2}Q\left( {\frac{{\gamma _{m,k}^{\rm{iq}}}}{{\sqrt {2\Delta _1^2} }}} \right)\notag\\
&- \frac{1}{2}Q\left( {\frac{{\gamma _{m,k}^{\rm{iq}} + {\vartheta_{m,k} ^{\rm{iq}}}}}{{\sqrt {\Delta _+^2 } }}} \right) + \frac{1}{2}Q\left( {\frac{{ - \gamma _{m,k}^{\rm{iq}} + {\vartheta_{m,k} ^{\rm{iq}}}}}{{\sqrt {\Delta _+^2} }}} \right). \tag{33}
\end{align}

{\it{Remark 4:}} We also present the near-optimal threshold and BER for the multi-channel multi-tag AmBC under the assumption of ideal TX and RX without IQ imbalance in (\ref{eq34}) and (\ref{eq35}). The results are consistent with those in \cite{wang2016ambient} for the single-channel single-tag AmBC under the assumption of ideal TX and RX without IQ imbalance. This is because, for an ideal direct-conversion wideband multi-channel receiver, the symbol detection on a given channel is the same as that for an ideal single-channel AmBC receiver. However, directly
employing these results will lead to performance degradation under IQ imbalance.
\begin{align}
\gamma _{m,k}^{\rm{ideal}} =& \frac{{\left| {\delta _{m,k}^{\rm{ideal}}} \right|}}{2} + \frac{{\varsigma _0^2 + \varsigma _1^2}}{{\left| {{\delta _{m,k}^{\rm{ideal}}}} \right|}}\notag \\
&\times \ln \left( {1 + \sqrt {1 - \exp \left( - \frac{{{{\left( {\delta _{m,k}^{\rm{ideal}}} \right)}^2}}}{{\varsigma _0^2 + \varsigma _1^2}} \right)} } \right),
\tag{34}
\label{eq34}
\end{align}
\begin{align}\label{eq35}
\mathbb{P}_{{\rm{BER}}}^{{\rm{ideal}}} =& \frac{1}{2}Q\left( {\frac{{\gamma _{m,k}^{\rm{ideal}}}}{{\sqrt {2\varsigma _0^2} }}} \right) + \frac{1}{2}Q\left( {\frac{{\gamma _{m,k}^{\rm{ideal}}}}{{\sqrt {2\varsigma _1^2} }}} \right)\notag\\
&- \frac{1}{2}Q\left( {\frac{{ \gamma _{m,k}^{\rm{ideal}} + {\delta_{m,k} ^{\rm{ideal}}}}}{{\sqrt {\varsigma _0^2 + \varsigma _1^2} }}} \right)+ \frac{1}{2}Q\left( {\frac{{-\gamma _{m,k}^{\rm{ideal}} + {\delta_{m,k} ^{\rm{ideal}}}}}{{\sqrt {\varsigma _0^2 + \varsigma _1^2} }}} \right),\tag{35}
\end{align}
where the parameters of $\delta _{m,k}^{\rm{ideal}}$, $\varsigma _0^2$ and $\varsigma _1^2$ can be respectively given by,
\begin{align}\label{eq36}
\delta _{m,k}^{\rm{ideal}} &= \mathbb{E}\left( {\Gamma _{m,k}^{{\rm{ideal}}}\left|B_m(k) = 1 \right.} \right) - \mathbb{E}\left( {\Gamma _{m,k}^{{\rm{ideal}}}\left| B_m(k) = 0 \right.} \right)\notag\\
&=\left( {{{\left| {{h_m} + {\mu _m}{g_m}} \right|}^2}{P_s} + \sigma {}_w^2} \right) - \left( {{{\left| {{h_m}} \right|}^2}{P_s} + \sigma {}_w^2} \right)\notag\\
&= \left( {{{\left| {{h_m} + {\mu _m}{g_m}} \right|}^2} - {{\left| {{h_m}} \right|}^2}} \right){P_s},\tag{36}
\end{align}
\begin{equation}
\begin{array}{l}
\varsigma _0^2 = {\mathbb{D}} \left( {\Gamma _{m,k}^{{\rm{ideal}}}\left| {B_m(k) = 0} \right.} \right)\\
{\quad} = \frac{2}{N}{\left| {{h_m}} \right|^2}{P_s}\sigma _w^2,
\end{array}
\tag{37}
\label{eq37}
\end{equation}
\begin{equation}
\begin{array}{l}
\varsigma _1^2 = {\mathbb{D}} \left( {\Gamma _{m,k}^{{\rm{ideal}}}\left| {B_m(k) = 1} \right.} \right)\\
{\quad} = \frac{2}{N}{\left| {{h_m} + {\mu_m}{g_m}} \right|^2}{P_s}\sigma _w^2.
\end{array}
\tag{38}
\label{eq38}
\end{equation}

\subsection{Blind Estimation of Parameters}

It is important to note that $\left|\vartheta _{m,k}^{{\rm{iq}}}\right|$ and $\Delta_+$ in (\ref{eq32}) are needed for the RX to calculate the threshold $\gamma _{m,k}^{\rm{iq}}$. However, obtaining the values of these two parameters requires CSI, IQ imbalance parameters, the presence probability of the incident signal of the image channel and backscattered signal of BT $-m$, noise power, and the transmit power of RX, all of which are unknown to RX. Thus, it is necessary to design a practical method to estimate these two parameters, as discussed below.

In the case of large $K$, $\mathbb{E}\left( {\left| {T_{m,k}^{\rm{iq}}} \right|} \right)$ and $\mathbb{D}\left( {{T_{m,k}^{\rm{iq}}}} \right)$ can be estimated at RX via (\ref{eq39}) and (\ref{eq40}), respectively, after receiving a sequence of symbols. It is worth mentioning that the choice of $K$ will be discussed in detail through simulations in the next section.
\begin{equation}
\mathbb{E}\left( {\left| {T_{m,k}^{{\rm{iq}}}} \right|} \right) = \frac{1}{K}\sum\limits_{k = 1}^K {\left| {T_{m,k}^{{\rm{iq}}}} \right|},
\tag{39}
\label{eq39}
\end{equation}
\begin{equation}
\mathbb{D}\left( { {T_{m,k}^{\rm{iq}}}} \right) = \frac{1}{{K - 1}}\sum\limits_{k = 1}^K {{{\left| {{T_{m,k}^{\rm{iq}}} - \mathbb{E}\left( { {T_{m,k}^{\rm{iq}}} } \right)} \right|}^2}} .
\tag{40}
\label{eq40}
\end{equation}

\emph{\textit{Lemma 4:}} $\left|\vartheta _{m,k}^{{\rm{iq}}}\right|$ and $\Delta_+$ can be approximated by solving (\ref{eq38}) and (\ref{eq39}).
\begin{align}\label{eq41}
\mathbb{E}\left( {\left| {T_{m,k}^{\rm{iq}}} \right|} \right)  \approx &\sqrt {\frac{3}{{8\pi }}} \Delta _{ + }^{} + \frac{{\Delta _{ + }^{}}}{{\sqrt {2\pi } }}\exp \left( { - \frac{{{{\left( {\vartheta _{m,k}^{{\rm{iq}}}} \right)}^2}}}{{2\Delta _{ + }^2}}} \right)\notag\\
&+ \frac{{\vartheta _{m,k}^{{\rm{iq}}}}}{{\sqrt {2\pi } }}\int_0^{\frac{{\vartheta _{m,k}^{{\rm{iq}}}}}{{\Delta _{ + }^{}}}} {{\exp \left({ - \frac{{{t^2}}}{2}}\right)}dt},\tag{41}
\end{align}
\begin{equation}
\mathbb{D}\left( {{T_{m,k}^{\rm{iq}}}} \right) = \Delta _{ + }^2 + \frac{1}{2}{\left( {\vartheta _{m,k}^{{\rm{iq}}}} \right)^2}.
\tag{42}
\label{eq42}
\end{equation}

\emph{Proof:} The detailed proof is presented in Appendix C.\hfill {$\blacksquare $}

\floatname{algorithm}{Algorithm}
\renewcommand{\thealgorithm}{1}
\begin{algorithm}[!t]
\setstretch{1}
   \caption{Bisection method for $\Delta^*$}
    \label{alg:AOA}
    \renewcommand{\algorithmicrequire}{\textbf{Input:}}
    \renewcommand{\algorithmicensure}{\textbf{Output:}}
    \begin{algorithmic}[1]
         \STATE {$\textbf{Inputting:}$ \\
         $\varepsilon_0$, $\varepsilon_1$, ${x_0}$, ${x_1}$ and $f\left(  \cdot  \right)$.
         \STATE $\textbf{Initialization:}$ \\
         Select ${\varepsilon _0} = {\varepsilon _1} = {10^{ - 3}}$, ${x_0}$=0, ${x_1} \le \sqrt {{\mathbb{D}\left( {\left| {T_{m,k}^{\rm{iq}}} \right|} \right)}}$ such that $f(x_0)$ and $f(x_1)$ have the opposite signs.
         \STATE $\textbf{Repetition:}$\\
         \STATE ${x_2} = \frac{{{x_0} + {x_1}}}{2}$, $f_0=f(x_0)$, $f_1=f(x_1)$, $f_2=f(x_2)$.
         \STATE Update $\left\{ {\begin{array}{*{20}{l}}
{{x_0}={x_0},\ {x_1}={x_2},\ if\ f_0 \cdot f_2 < 0}\\
{{x_0}={x_2},\ {x_1}={x_1},\ if\ f_2 \cdot f_1 < 0}
\end{array}} \right..$
         \STATE Repeat steps 4 and 5 until $\left| {{x_1} - {x_0}} \right| < {\varepsilon _0}$ or  $\left| {f\left( {{x_2}} \right)} \right| < {\varepsilon _1}$ or $f(x_2)=0$.

         \STATE $\textbf{Output:}$ The satisfying value $\Delta^* = x_2$.}
    \end{algorithmic}
    \label{AA1}
\end{algorithm}\

To obtain the value of $\Delta_+$, we first formulate the function $f(\Delta_+)$ based on (\ref{eq41}) and (\ref{eq42}) as
\begin{align}\label{eq43}
f\left( {{\Delta _ + }} \right) = &\sqrt {\frac{3}{{8\pi }}} {\Delta _ + } + \frac{{{\Delta _ + }}}{{\sqrt {2\pi } }}\exp \left( { - \frac{{\mathbb{D}\left( { {T_{m,k}^{\rm{iq}}} } \right) - \Delta _ + ^2}}{{\Delta _ + ^2}}} \right)\notag\\
&+ \sqrt {\frac{{\left( {\mathbb{D}\left( {{T_{m,k}^{\rm{iq}}} } \right) - \Delta _ + ^2} \right)}}{\pi }}\notag\\
& \times{\int_0^{\frac{{\sqrt {2\left( {\mathbb{D}\left( { {T_{m,k}^{\rm{iq}}} } \right) - \Delta_+ ^2} \right)} }}{{{\Delta _ + }}}} {{\exp \left({ - \frac{{{t^2}}}{2}}\right)}dt}}- \mathbb{E}\left( {\left| {T_{m,k}^{\rm{iq}}} \right|} \right),\tag{43}
\end{align}
where $f(\Delta_+)=0$. In order to solve this equation, we utilize the one-dimensional search method to approximate the solution. By following the Algorithm 1, the approximated optimal parameter value of $\Delta_+$ can be given. Then, by substituting the value of $\Delta_+$ into (\ref{eq42}), the value of $\left|\vartheta _{m,k}^{{\rm{iq}}}\right|$ can be obtained.

{\it{Remark 5:}} We call this estimation method a blind method, where $K$ unknown symbols are utilized to blindly estimate values of $\left|\vartheta _{m,k}^{{\rm{iq}}}\right|$ and $\Delta_+$ without requiring pilot symbols.

\section{Numerical Results And Discussions}
In this section, we will conduct numerical analyses to verify the derived results. We will also study the impacts of IQ imbalance parameters and interference parameters. Unless otherwise specified, the parameters used in this paper are as follows. The channels $h_m$, $\mu _m$, $g_m$, $h_{-m}$, $\mu _{-m}$ and $g_{-m}$ are complex Gaussian random variables with zero mean and unit variance. The noise variance is set to 1, the sample size is set to $N=100$, and the SNR is set to 15 $\rm{dB}$. The TX signal presence probability at the image channel and the active probability at the corresponding BT are both set to 0.5, i.e., $q=v=0.5$.

\begin{figure}
  \centering
  {\includegraphics[width=0.45\textwidth]{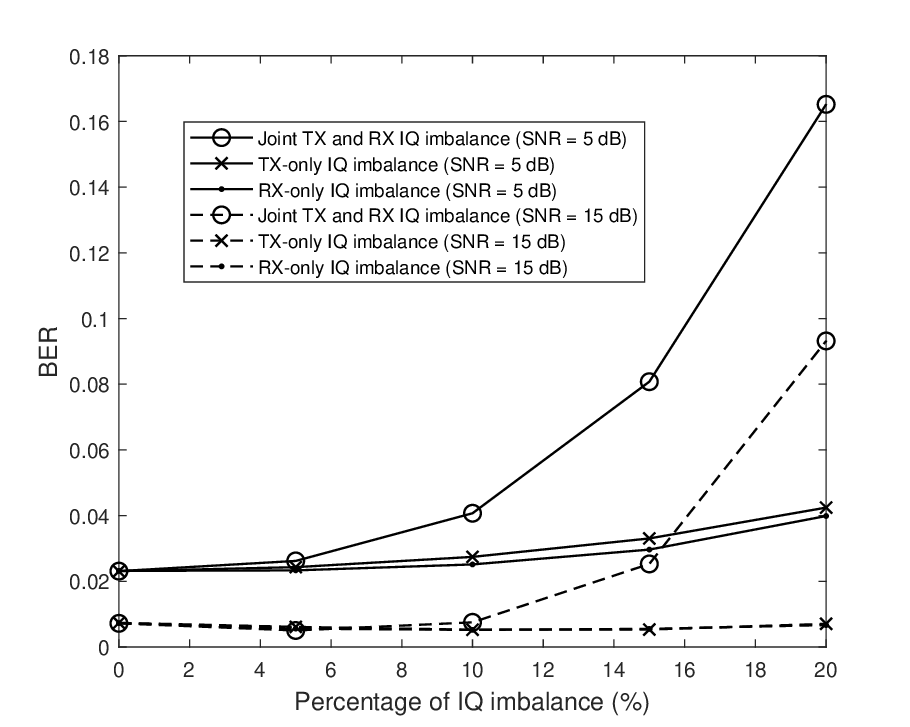}}\\
  \caption{BER versus the percentage of IQ imbalance using the detection threshold $\gamma _{m,k}^{{\rm{ideal}}}$ assuming ideal transceiver when SNR $=5\ \rm{dB}$ and SNR $=15\ \rm{dB}$.}\label{fig4}
\end{figure}

The impact of different levels of IQ imbalance on BER performance are illustrated in Fig. \ref{fig4}, with the detection threshold $\gamma _{m,k}^{{\rm{ideal}}}$ in (\ref{eq34}) assuming ideal transceiver. The SNR values are set to $5\ \rm{dB}$ and $15\ \rm{dB}$. The percentage of IQ imbalance is defined by the deviations of amplitude and phase from their ideal values \footnote{According to \cite{6746245}, practical IQ imbalance levels in transceivers with high-quality RF circuit design typically range from 1\% to 5\% for both amplitude and phase, implying that IQ imbalance is inevitable and may be more pronounced in cost-constrained AmBC systems.}. The amplitude deviation is given by $(1-\rho)\times 100\%$, where $\rho^t = \rho^r = 0.95$ and $0.9$ correspond to 5\% and 10\%, respectively. The phase deviation is given by $\tfrac{2\phi}{\pi}\times 100\%$, where $\phi^t = \phi^r = \pi/36$ and $\pi/18$ correspond to 5.56\% and 11.11\%. Unless stated otherwise, the amplitude and phase deviations are assumed to change simultaneously at the same percentage for ease of presentation. It can be observed that when SNR $= 5\ \rm{dB}$, BER increases with the percentage IQ imbalance, indicating that IQ imbalance significantly degrades the symbol detection performance with $\gamma _{m,k}^{{\rm{ideal}}}$. Additionally, the performance degradation caused by TX IQ imbalance is much higher than that caused by RX IQ imbalance. Furthermore, the joint TX and RX IQ imbalance results in the most severe degradation, particularly at high levels of IQ imbalance. This indicates that IQ imbalance can not be neglected in AmBC and should be carefully considered. It can also be observed that when SNR $= 15\ \rm{dB}$, an IQ imbalance exceeding $10\%$ leads to a noticeable performance degradation due to the joint TX and RX IQ imbalance, whereas the impact of TX-only or RX-only imbalance remains insignificant.

\begin{table*}[t]
\centering
\caption{Comparison of Detection Thresholds}
\begin{tabular}{|c|c|c|c|}
\hline
IQ imbalance parameters and probability & \makecell{$\rho^t = \rho^r = 1$, \\ $\phi^t = \phi^r = 0$ \\ and $q=0, v=0$} & \makecell{$\rho^t = \rho^r = 0.95$, \\ $\phi^t = \phi^r = \frac{\pi}{36}$ \\ and $q=0.5, v=0.5$} & \makecell{$\rho^t = \rho^r = 0.9$,\\ $\phi^t = \phi^r = \frac{\pi}{18}$ \\ and $q=0.5, v=0.5$}

\\ \hline
{Our theoretical $ {\gamma_{m,k}^{\rm{iq}}} $} & 11.5192 & 10.3447 & 9.1459 \\ \hline
{Our estimated $ {\gamma_{m,k}^{\rm{iq}}} $} & 11.5722 & 10.3648
 &  9.2102 \\ \hline
{the optimal $ {\gamma_{m,k}^{\rm{iq}}} $ (Fig. \ref{fig5})} & 12 & 10.9 & 9.95 \\ \hline

\end{tabular}
\label{table1}
\end{table*}

\begin{figure}
  \centering
  \includegraphics[width=0.45\textwidth]{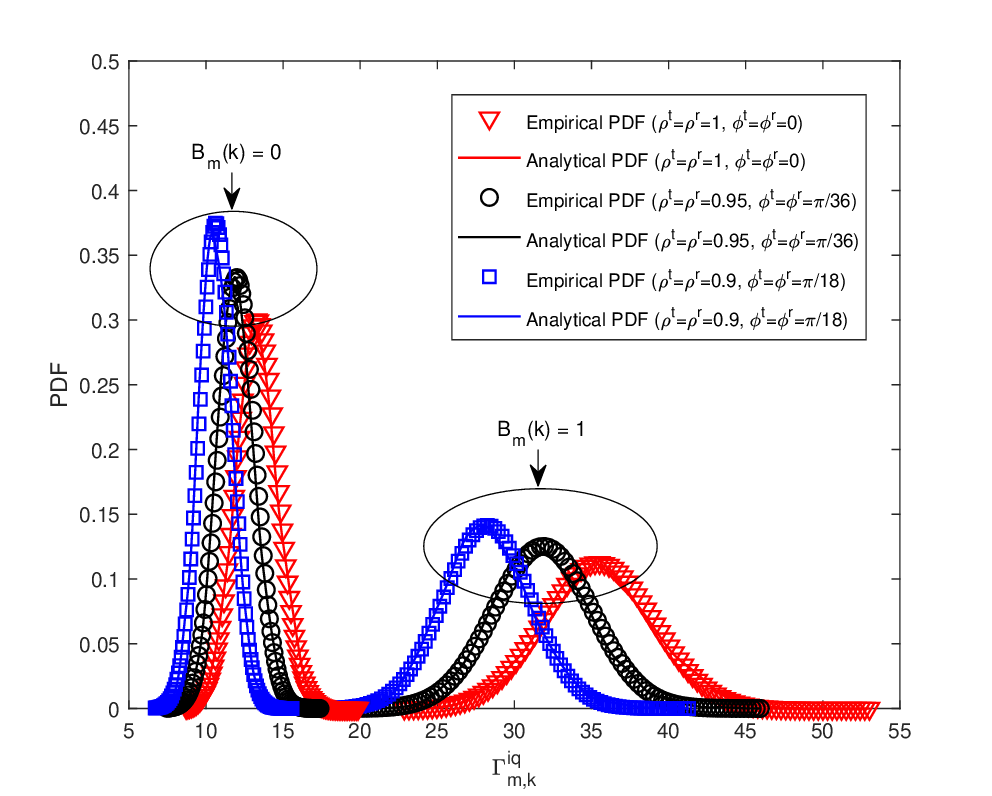}\\
  \caption{Empirical and analytical PDFs for $\Gamma _{m,k}^{{\rm{iq}}}$.}\label{fig5}
\end{figure}

Fig. \ref{fig5} shows the comparison of the empirical and theoretical PDF of $\Gamma _{m,k}^{\rm{iq}}$ when $\rho^r=\rho^t=1$ and $\varphi^r=\varphi^t=0$, $\rho^r=\rho^t=0.95$ and $\varphi^r=\varphi^t=\frac{\pi}{36}$, $\rho^r=\rho^t=0.9$ and $\varphi^r=\varphi^t=\frac{\pi}{18}$, respectively. We can see that the empirical PDFs of $\Gamma _{m,k}^{\rm{iq}}$ closely match the analytical PDFs of $\Gamma _{m,k}^{\rm{iq}}$. We can also see that the PDFs shift to the left as the levels of both amplitude and phase imbalance increase. This shift occurs because the increase in IQ imbalance levels can cause energy leak into the image channel, leading to the decrease of $\Gamma _{m,k}^{{\rm{iq}}}$.

\begin{figure}
  \centering
  \includegraphics[width=0.45\textwidth]{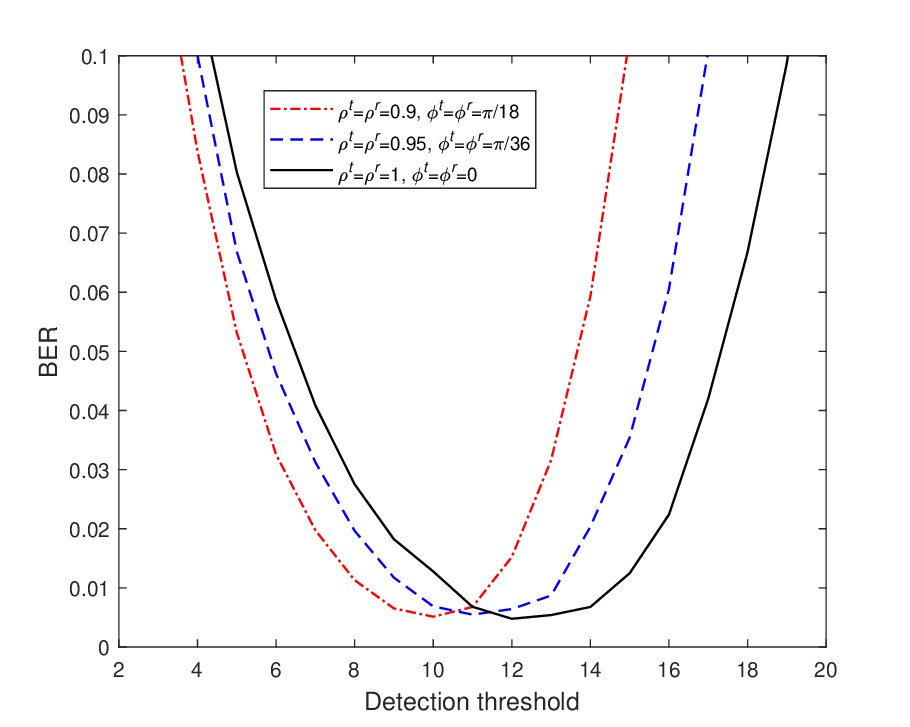}\\
  \caption{BER versus detection threshold with detection threshold $\gamma _{m,k}^{{\rm{iq}}}$ when SNR $=5\ \rm{dB}$ and SNR $=15\ \rm{dB}$.}\label{fig6}
\end{figure}

 Fig. \ref{fig6} shows the relationship between BER and detection threshold under IQ imbalance. As can be seen, for different IQ imbalance parameters, BER decreases as the detection threshold increases, reaches a minimum point, and then rises, which indicates that an optimal detection threshold exists by minimizing BER. In addition, the optimal detection threshold decreases when the levels of IQ imbalance increases. In Table \uppercase\expandafter{\romannumeral 1}, we present the optimal detection thresholds from Fig. \ref{fig6}, the theoretical near-optimal detection thresholds calculated by (\ref{eq32}), and the estimated near-optimal detection thresholds obtained based on Algorithm 1. We can see that our theoretical and estimated detection thresholds are both close to the optimal detection threshold.
\begin{figure}
  \centering
  \includegraphics[width=0.45\textwidth]{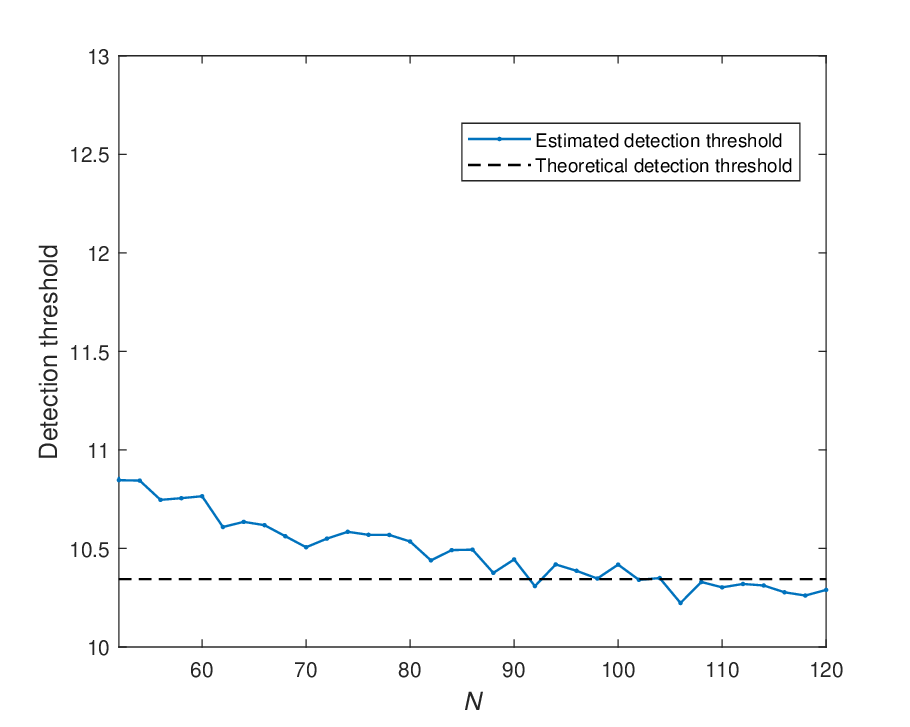}\\
  \caption{Detection threshold versus sample size $N$.}\label{fig7}
\end{figure}

\begin{figure}
  \centering
  \includegraphics[width=0.45\textwidth]{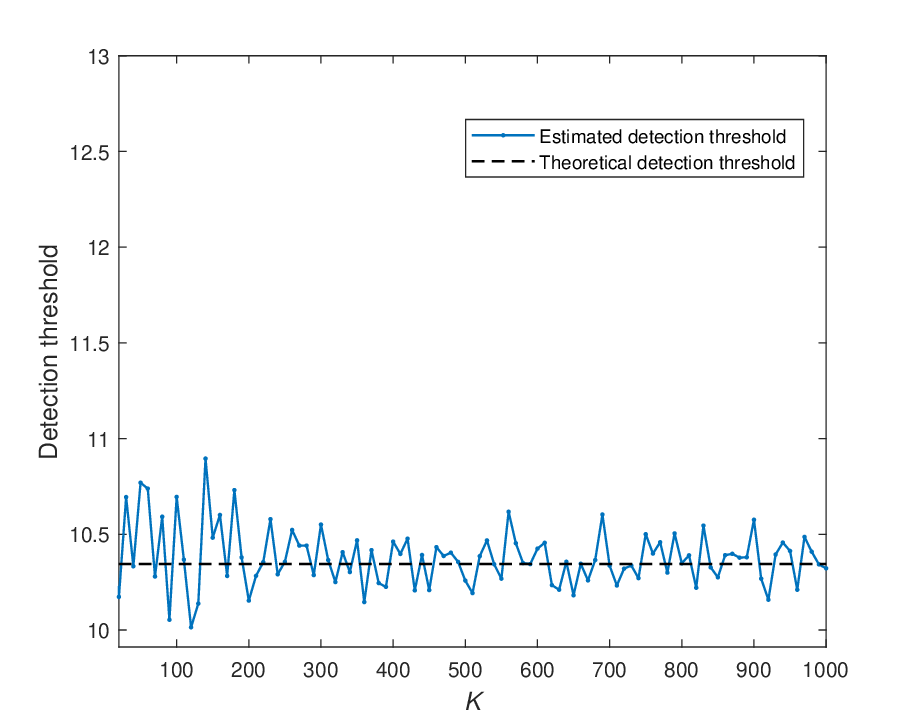}\\
  \caption{Detection threshold versus symbol number $K$.}\label{fig8}
\end{figure}

To further study the impacts of sample size $N$ and symbol number $K$ on the estimated detection thresholds under IQ imbalance, Fig. \ref{fig7} and Fig. \ref{fig8} are plotted. Fig. \ref{fig7} illustrates that as the sample size $N$ increases, the estimated threshold becomes closer with the theoretical near-optimal threshold, which indicates that larger sample sizes improves the accuracy of the estimated threshold. Fig. \ref{fig8} shows that as the symbol number $K$ increases, the estimated threshold becomes closer with the theoretical near-optimal threshold. Moreover, for all values of $K$, the estimated threshold remains consistently close to the theoretical one, indicating that the estimated near-optimal threshold can be obtained with relatively few symbols.

\begin{figure}
  \centering
  \includegraphics[width=0.45\textwidth]{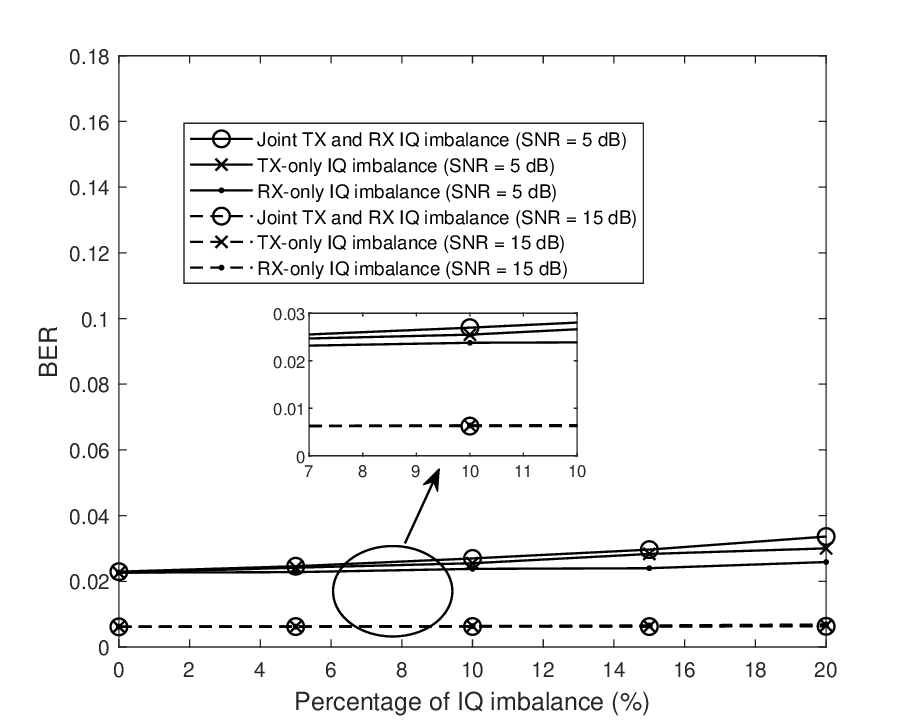}\\
  \caption{BER versus the percentage of IQ imbalance using the detection threshold $\gamma _{m,k}^{{\rm{iq}}}$ when SNR $=5\ \rm{dB}$ and SNR $=15\ \rm{dB}$.}\label{fig9}
\end{figure}

In Fig. \ref{fig9}, we analyze the impact of IQ imbalance on the BER performance when using the detection threshold $\gamma _{m,k}^{\rm{iq}}$ derived in (\ref{eq32}). For ease of comparison, the parameter settings are the same as those in Fig. \ref{fig4}. From Fig. \ref{fig9}, one can see that the BER slightly increases as the percentage of IQ imbalance increases when SNR $= 5\ \rm{dB}$,  whereas at SNR $= 15\ \rm{dB}$, it remains nearly unchanged. By comparing Fig. Fig. \ref{fig9} and Fig. Fig. \ref{fig4}, we can see that the performance degradation caused by IQ imbalance has been significantly reduced when SNR $ = 5\  \rm{dB}$. For example, when the percentage of IQ imbalance is $10\%$, the BER under joint TX and RX IQ imbalance, TX-only IQ imbalance, and RX-only IQ imbalance based on the detection threshold $\gamma _{m,k}^{\rm{ideal}}$, are 0.0407, 0.0274, 0.0251, respectively, as shown in Fig. \ref{fig4}. In contrast, under the same conditions but using the detection threshold $\gamma _{m,k}^{\rm{iq}}$, the BER reduce to 0.0269, 0.0254, 0.0237, respectively, as shown in Fig. 9. Furthermore, when  SNR $ = 15 \ \rm{dB}$, the performance degradation caused by joint TX and RX IQ imbalance can also be alleviated. For instance, when the IQ imbalance percentage is $10\%$, the BER under joint TX and RX IQ imbalance, based on the detection thresholds $\gamma _{m,k}^{\rm{ideal}}$ and $\gamma _{m,k}^{\rm{iq}}$, are 0.0075 and 0.0064, respectively.

\begin{figure}
  \centering
  \includegraphics[width=0.45\textwidth]{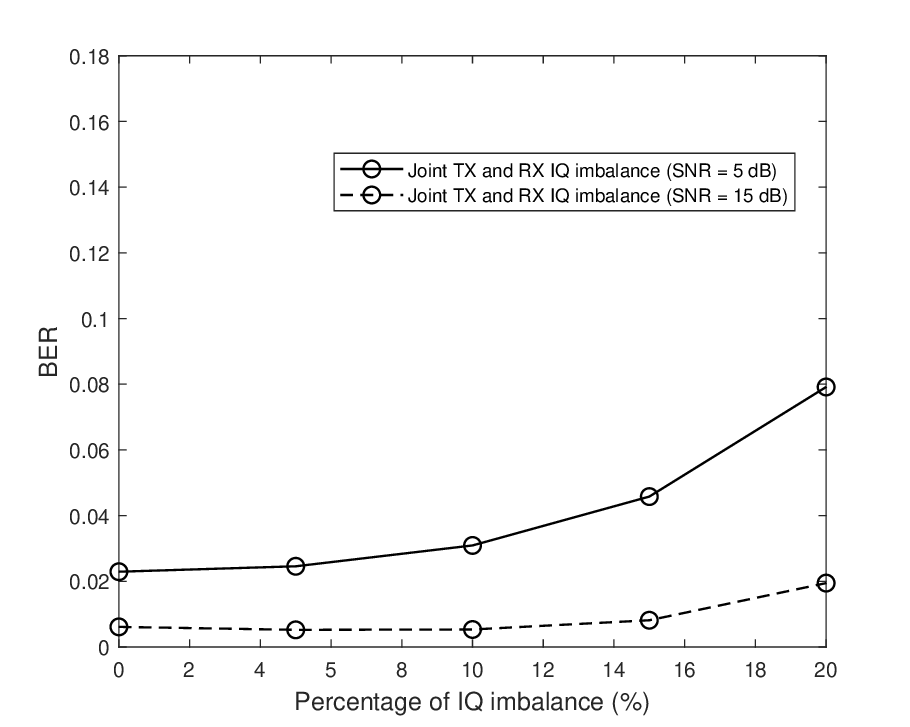}\\
  \caption{BER versus the percentage of IQ imbalance using the detection threshold $\gamma^{\rm{iq-RX}}$ given in \cite{10005249} assuming non-ideal RX when SNR $=5\ \rm{dB}$ and SNR $=15\ \rm{dB}$.}\label{fig10}
\end{figure}

In Fig. \ref{fig10}, we present the impact of IQ imbalance on the BER performance when using the detection threshold $\gamma^{\rm{iq-RX}}$ [24, eq.(10)], assuming non-ideal RX with RX-only IQ imbalance. For ease of comparison, the parameter settings are the same as those in Fig. \ref{fig4} and Fig. \ref{fig9}. By comparing Fig. \ref{fig10} and Fig. \ref{fig4}, it can be observed that using the detection threshold $\gamma^{\rm{iq-RX}}$ alleviates BER performance degradation under joint TX and RX imbalance. However, by comparing Fig. \ref{fig10} and Fig. \ref{fig9}, it can be seen that using the proposed $\gamma^{\rm{iq}}_{m,k}$ outperforms using the detection threshold $\gamma^{\rm{iq-RX}}$ under joint TX and RX IQ imbalance.

\begin{figure}
  \centering
  \includegraphics[width=0.45\textwidth]{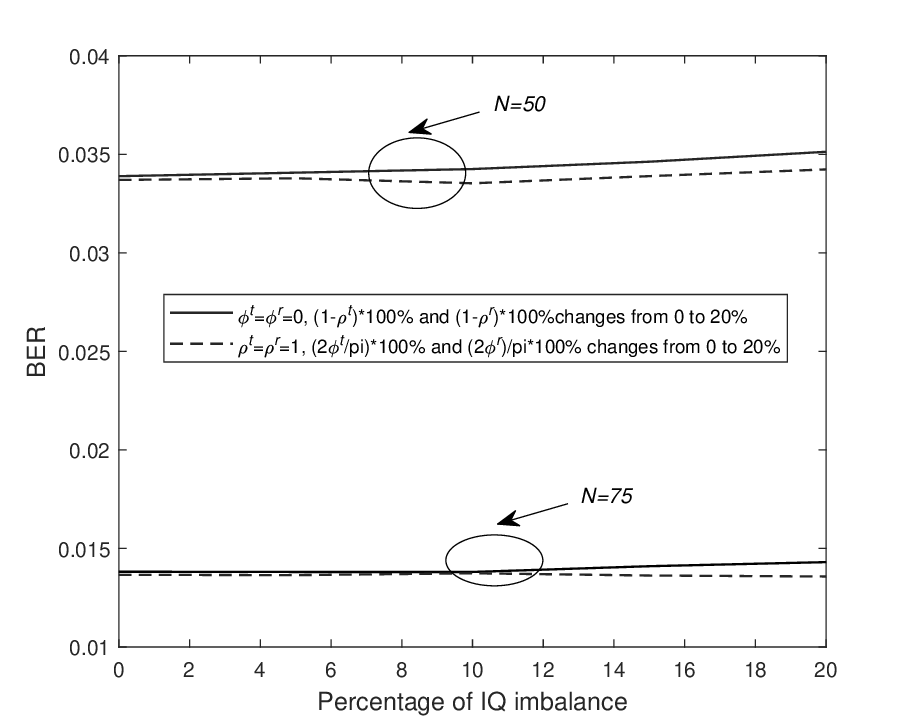}\\
  \caption{BER versus the parameters of amplitude and phase imbalance for $N=50$ and $N=75$ with detection threshold $\gamma _{m,k}^{{\rm{iq}}}$.}\label{fig11}
\end{figure}

In order to elaborate the impacts of IQ imbalance parameters on the BER with the derived detection threshold $\gamma _{m,k}^{\rm{iq}}$. Fig. \ref{fig11} illustrates how BER changes with the parameters of amplitude and phase imbalance, ranging from 0 to 20\%, for $N=50$ and $N=75$, respectively. It can be observed that the BER increases as the percentage of IQ imbalance increases. It can also be seen that the amplitude imbalance leads to a greater performance degradation compared to the phase imbalance, which indicates that the detector based on our detection threshold is more sensitive to the amplitude imbalance than to the phase imbalance.

\begin{figure}
  \centering
  \includegraphics[width=0.45\textwidth]{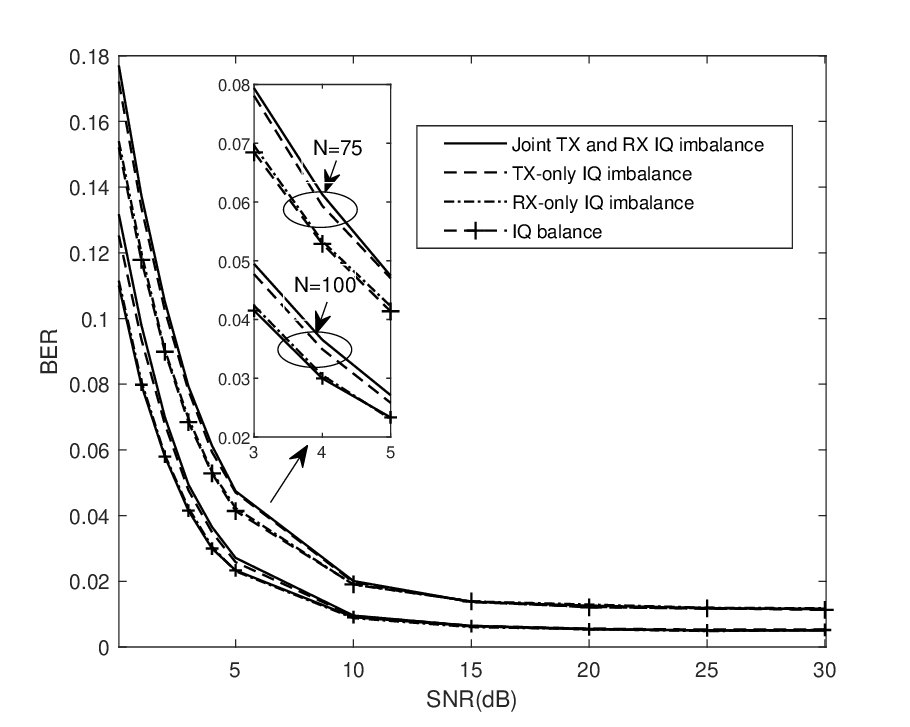}\\
  \caption{BER versus SNRs for $N=75,100$ with detection threshold $\gamma _{m,k}^{{\rm{iq}}}$.}\label{fig12}
\end{figure}

Fig. \ref{fig12} presents the relationship between BER and SNR based on our derived detection threshold $\gamma _{m,k}^{\rm{iq}}$ under joint TX and RX IQ imbalance with $\rho^t = \rho^r = 0.9$, $\phi^t = \phi^r = \frac{\pi}{18}$, TX-only IQ imbalance with $\rho^t = 0.9$, $\rho^r = 0$, $\phi^t = \frac{\pi}{18}$, $\phi^r = 0$, RX-only IQ imbalance with $\phi^t = 0$, $\phi^r = \frac{\pi}{18}$, $\rho^t = 1$, $\rho^r = 0.9$ and IQ balance with $\rho^t = \rho^r = 1$, $\phi^t = \phi^r = 0$. The sample size is set to $N=75$ and $N=100$. It can be observed that the performance degradation caused by IQ imbalance is trivial when using our derived  detection threshold $\gamma _{m,k}^{\rm{iq}}$. It can also be observed that an error floor exists that is caused by both direct link interference and IQ imbalance interference.

\begin{figure}
  \centering
  \includegraphics[width=0.45\textwidth]{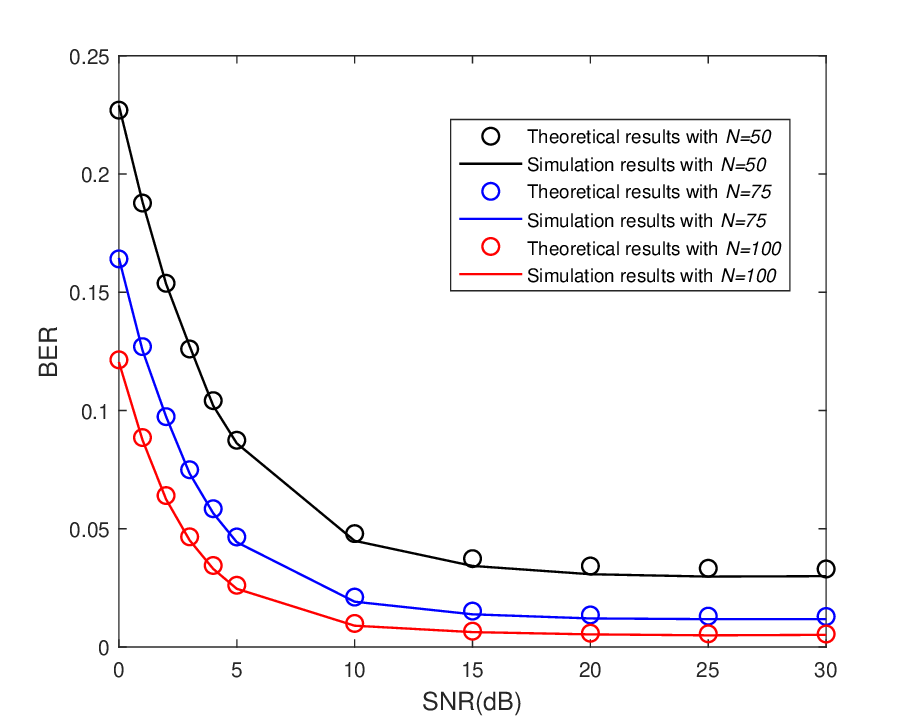}\\
  \caption{Theoretical and simulation BER versus SNRs for $N=50$, $N=75$, and $N=100$ with detection threshold $\gamma _{m,k}^{{\rm{iq}}}$.}\label{fig13}
\end{figure}

\begin{figure}
  \centering
  \includegraphics[width=0.45\textwidth]{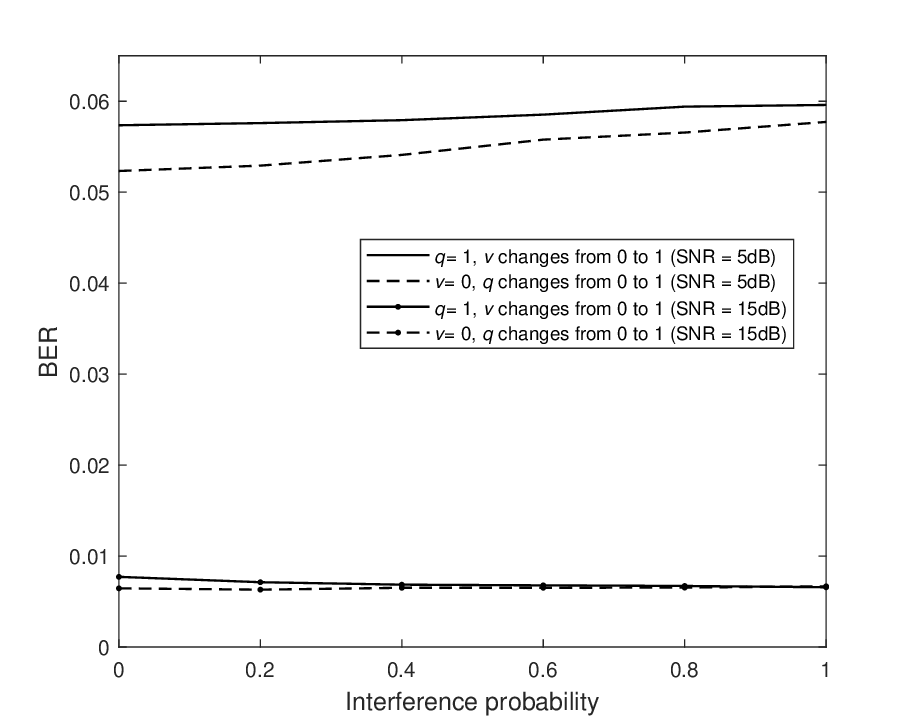}\\
  \caption{BER versus interference probabilities for indicated IQ imbalance values when SNR $=5\ \rm{dB}$ and SNR $=15\ \rm{dB}$ .}\label{fig14}
\end{figure}

Fig. \ref{fig13} shows the BER versus SNR under joint TX and RX IQ imbalance with $\rho^t = \rho^r = 0.95$, $\phi^t = \phi^r = \frac{\pi}{36}$ for different sample numbers. The sample size is set to $N=50$, $N=75$, and $N=100$. Clearly, the theoretical BER closely matches the simulated one and the larger $N$ results in the lower BER. Fig. \ref{fig14} illustrates BER versus interference probabilities under IQ imbalance when using the derived detection threshold $\gamma _{m,k}^{\rm{iq}}$. The SNR values are set to $5\ \rm{dB}$ and $15\ \rm{dB}$, and the IQ imbalance parameters are set to $\rho^t = \rho^r = 0.6$ and $\phi^t = \phi^r =\frac{\pi}{10}$. It can be seen that BER increases with the increase of the signal presence probability at the image channel and the active probability at the corresponding BT when SNR is 5$\rm{dB}$, and remains nearly unchanged despite variations in the probabilities when SNR is 15$\rm{dB}$.


\begin{figure*}[!b]
\hrulefill
\[\begin{array}{l}
\Gamma _{m,k\left| {\xi,\eta} \right.}^{{\rm{iq}}}\\
 = \left\{ \begin{array}{l}
\frac{1}{N}\sum\limits_{n = (k - 1)N + 1}^{kN} \begin{array}{l}
{\left| {\left( {{I_1}{h_m} + {I_2}} \right){p_m}\left( n \right)} \right|^2} + {\left| {\left( {{I_3}{h_m} + {I_4}} \right)\xi {p_{ - m}}^*\left( n \right)} \right|^2} + {\left| {{\tau _w}(n)} \right|^2}\\
 + 2\left| {\left( {{I_1} + {I_2}} \right){p_m}\left( n \right)} \right|\left| {\left( {{I_3} + {I_4}} \right)\xi {p_{ - m}}^*\left( n \right)} \right|\\
 + 2\left| {\left( {{I_1} + {I_2}} \right){p_m}\left( n \right)} \right|\left| {{\tau _w}(n)} \right| + 2\left| {\left( {{I_3} + {I_4}} \right)\xi {p_{ - m}}^*\left( n \right)} \right|\left| {{\tau _w}(n)} \right|,\ B_m(k) = 0
\end{array} \\
\frac{1}{N}\sum\limits_{n = (k - 1)N + 1}^{kN} \begin{array}{l}
{\left| {\left( {{I_1}\left( {{h_m} + {\mu _m}{g_m}} \right) + {I_2}} \right){p_m}\left( n \right)} \right|^2} + {\left| {\left( {{I_3}\left( {{h_m} + {\mu _m}{g_m}} \right) + {I_4}} \right)\xi {p_{ - m}}^*\left( n \right)} \right|^2}\\
 + {\left| {{\tau _w}(n)} \right|^2} + 2\left| {\left( {{I_1}\left( {{h_m} + {\mu _m}{g_m}} \right) + {I_2}} \right){p_m}\left( n \right)} \right|\left| {\left( {{I_3}\left( {{h_m} + {\mu _m}{g_m}} \right) + {I_4}} \right)\xi {p_{ - m}}^*\left( n \right)} \right|\\
 + 2\left| {\left( {{I_1}\left( {{h_m} + {\mu _m}{g_m}} \right) + {I_2}} \right){p_m}\left( n \right)} \right|\left| {{\tau _w}(n)} \right| + 2\left| {\left( {{I_3}\left( {{h_m} + {\mu _m}{g_m}} \right) + {I_4}} \right)\xi {p_{ - m}}^*\left( n \right)} \right|\left| {{\tau _w}(n)} \right|,\ B_m(k) = 1
\end{array}
\end{array} \right.
\tag{A.2}
\label{eqA.2}
\end{array}\]

\end{figure*}
\section{Conclusion}
In this paper, we proposed a symbol detection framework for multi-channel multi-tag AmBC network by considering the ambient RF source and AmBC receiver impaired by IQ imbalance. Considering differential encoding and energy difference detector, we derived the BER and near-optimal detection thresholds. We also devised a threshold estimation method using signal samples at the AmBC receiver, which did not require any prior information. Finally, simulation results verified the derived results and demonstrated that the performance degradation caused by IQ imbalance can be significantly reduced by using our derived detection threshold.

\section*{Appendix A}
Let the RX noise impaired by IQ imbalance be $\tau _w(n)$ that is given by
\begin{equation}
{\tau _w}(n) = \kappa _1^r{w_m}(n) + \kappa _2^r{w_{-m}}^ * (n).
\tag{A.1}
\label{eqA.1}
\end{equation}
Then, $\Gamma _{m,k\left| {\xi,\eta} \right.}^{{\rm{iq}}}$ can be expressed as (A.2), where
\begin{equation}
{I_1} = \kappa _1^r\kappa _1^t,
\tag{A.3}
\label{eqA.3}
\end{equation}
\begin{equation}
{I_2} = \kappa _2^r{\left( {\kappa _2^t} \right)^ * }{\left( {{h_{ - m}} + \eta {\mu _{ - m}}{g_{ - m}}} \right)^ * },
\tag{A.4}
\label{eqA.4}
\end{equation}
\begin{equation}
{I_3} = \kappa _1^r\kappa _2^t,
\tag{A.5}
\label{eqA.5}
\end{equation}
\begin{equation}
{I_4} = \kappa _2^r{\left( {\kappa _1^t} \right)^ * }{\left( {{h_{ - m}} + \eta {\mu _{ - m}}{g_{ - m}}} \right)^ * }.
\tag{A.6}
\label{eqA.6}
\end{equation}
\begin{equation}
{I_5} = \left( {{{\left| {\kappa _1^r} \right|}^2} + {{\left| {\kappa _2^r} \right|}^2}} \right).
\tag{A.7}
\label{eqA.7}
\end{equation}
According to central limit theorem (CLT) \cite{papoulis2002probability}, when $N$ is large, ${\Gamma _{m,k\left| {\xi ,\eta } \right.}^{{\rm{iq}}}}$ can be approximated as a Gaussian variable, with its mean and variance given by,
\begin{equation}
\left\{ \begin{array}{l}
\mathbb{E}\left( {\Gamma _{m,k\left| {\xi ,\eta } \right.}^{{\rm{iq}}}\left| B_m(k) = 0 \right.} \right)\\
 = {\left| {{I_1}{h_m} + {I_2}} \right|^2}{P_s} + {\left| {{I_3}{h_m} + {I_4}} \right|^2}\xi {P_s} + {I_5}\sigma _w^2\\
\mathbb{E}\left( {\Gamma _{m,k\left| {\xi ,\eta } \right.}^{{\rm{iq}}}\left| B_m(k) = 1 \right.} \right)\\
 = {\left| {{I_1}\left( {{h_m} + {\mu _m}{g_m}} \right) + {I_2}} \right|^2}{P_s}\\
 \quad + {\left| {{I_3}\left( {{h_m} + {\mu _m}{g_m}} \right) + {I_4}} \right|^2}\xi {P_s} + {I_5}\sigma _w^2
\end{array} \right.
\tag{A.8}
\label{eqA.8}
\end{equation}
and
\begin{equation}
\left\{ {\begin{array}{*{20}{l}}
{\mathbb{D}\left( {\Gamma _{m,k\left| {\xi,\eta} \right.}^{{\rm{iq}}}\left| B_m(k) = 0 \right.} \right)}\\
{ = \frac{1}{N}{{\left( {{{\left| {{I_1}{h_m} + {I_2}} \right|}^2}{P_s} + {{\left| {{I_3}{h_m} + {I_4}} \right|}^2}\xi {P_s} + {I_5}\sigma _w^2} \right)}^2}}\\
{\mathbb{D}\left( {\Gamma _{m,k\left| {\xi,\eta} \right.}^{{\rm{iq}}}\left| B_m(k) = 1 \right.} \right)}\\
{ = \frac{1}{N}{{\left( \begin{array}{l}
{\left| {{I_1}\left( {{h_m} + {\mu _m}{g_m}} \right) + {I_2}} \right|^2}{P_s}\\
 + {\left| {{I_3}\left( {{h_m} + {\mu _m}{g_m}} \right) + {I_4}} \right|^2}\xi {P_s} + {I_5}\sigma _w^2
\end{array} \right)}^2}}
\end{array}} \right..
\tag{A.9}
\label{eqA.9}
\end{equation}
Theorem 1 can be obtained by substituting (\ref{eqA.2})-(\ref{eqA.7})
into (\ref{eqA.8}) and (\ref{eqA.9}).

\begin{figure*}
\begin{align}\label{eqC.3}
\mathbb{E}\left( \left| T_{m,k}^{iq} \right| \right) &= \frac{1}{4}\sqrt {2\Delta _{0}^2} \sqrt {\frac{2}{\pi }}  + \frac{1}{4}\sqrt {2\Delta _{1}^2} \sqrt {\frac{2}{\pi }}  + \sqrt {\frac{{\Delta _{ + }^2}}{{8\pi }}} \exp \left( { - \frac{{{{\left( {\vartheta _{m,k}^{{\rm{iq}}}} \right)}^2}}}{{2\Delta _{ +}^2}}} \right) + \frac{1}{4}\vartheta _{m,k}^{{\rm{iq}}}\left( {1 - 2\Phi \left( { - \frac{{\vartheta _{m,k}^{{\rm{iq}}}}}{{\sqrt {\Delta _{ + }^2} }}} \right)} \right)\notag\\
&+ \sqrt {\frac{{\Delta _{ + }^2}}{{8\pi }}} \exp \left( { - \frac{{{{\left( {\vartheta _{m,k}^{{\rm{iq}}}} \right)}^2}}}{{2\Delta _{ + }^2}}} \right) - \frac{1}{4}\vartheta _{m,k}^{{\rm{iq}}}\left[ {1 - 2\Phi \left( { - \frac{{ - \vartheta _{m,k}^{{\rm{iq}}}}}{{\sqrt {\Delta _{ + }^2} }}} \right)} \right]\notag\\
= & \frac{\sqrt {\Delta _{0}^2}}{{2\sqrt {\pi } }}  + \frac{\sqrt {\Delta _{1}^2}}{{2\sqrt {\pi } }}  + \sqrt {\frac{{\Delta _{ + }^2}}{{2\pi }}} \exp \left( { - \frac{{{{\left( {\vartheta _{m,k}^{{\rm{iq}}}} \right)}^2}}}{{2\Delta _{ + }^2}}} \right) - \frac{1}{2}\vartheta _{m,k}^{{\rm{iq}}}\left( {\Phi \left( { - \frac{{\vartheta _{m,k}^{{\rm{iq}}}}}{{\sqrt {\Delta _{ + }^2} }}} \right) - \Phi \left( { - \frac{{ - \vartheta _{m,k}^{{\rm{iq}}}}}{{\sqrt {\Delta _{ + }^2} }}} \right)} \right)\notag\\
= & \frac{{\Delta _{0}^{} + \Delta _{1}^{}}}{{2\sqrt {\pi } }} + \frac{{\Delta _{ + }^{}}}{{\sqrt {2\pi } }}\exp \left( { - \frac{{{{\left( {\vartheta _{m,k}^{{\rm{iq}}}} \right)}^2}}}{{2\Delta _{ + }^2}}} \right) - \frac{1}{2}\vartheta _{m,k}^{{\rm{iq}}}\left( {\frac{1}{{\sqrt {2\pi } }}\int_{ - \infty }^{ - \frac{{\vartheta _{m,k}^{{\rm{iq}}}}}{{\Delta _{ + }^{}}}} {{e^{ - \frac{{{t^2}}}{2}}}dt}  - \frac{1}{{\sqrt {2\pi } }}\int_{ - \infty }^{\frac{{\vartheta _{m,k}^{{\rm{iq}}}}}{{\Delta _{ + }^{}}}} {{e^{ - \frac{{{t^2}}}{2}}}dt} } \right)\notag\\
\mathop  = \limits^{\Delta _{0}^{} + \Delta _{1}^{} \approx \sqrt {\frac{3}{2}} \Delta _{ + }^{}}& \sqrt {\frac{3}{{8\pi }}} \Delta _{ + }^{} + \frac{{\Delta _{ + }^{}}}{{\sqrt {2\pi } }}\exp \left( { - \frac{{{{\left( {\vartheta _{m,k}^{{\rm{iq}}}} \right)}^2}}}{{2\Delta _{ + }^2}}} \right) + \frac{{\vartheta _{m,k}^{{\rm{iq}}}}}{{\sqrt {2\pi } }}\int_0^{\frac{{\vartheta _{m,k}^{{\rm{iq}}}}}{{\Delta _{ + }^{}}}} {{e^{ - \frac{{{t^2}}}{2}}}dt}.\tag{C.3}
\end{align}
\hrulefill
\end{figure*}
\section*{Appendix B}
In most practical communication scenarios, the transmission probability for both information bit ``1'' and bit ``0'' are assumed to be equal, that is, ${\rm{Pr}}\left( {A_m(k) = 1} \right) = {\rm{Pr}}\left( {A_m(k) = 0} \right) = \frac{1}{2}$. Based on the assumption that BTs remain sleep state or enters active state with probability $\Pr \left( \rm{{state  = sleep}} \right) = 1-v$ and $\Pr \left( \rm{{state  = active}} \right) = v$, $0 \le v \le 1$, and the assumption of equiprobably transmitted bits, we have
\begin{align}\label{eqB.1}
&\Pr \left( {\eta  = 0} \right)\notag\\
= &\Pr \left( {A_m(k) = 0\left| \rm{{state = sleep}} \right.} \right)\Pr \left( \rm{{state = sleep}} \right)\notag\\
&+ \Pr \left( {A_m(k) = 1\left| \rm{{state = sleep}} \right.} \right)\Pr \left( \rm{{state = sleep}} \right)\notag\\
&+ \Pr \left( {A_m(k) = 0\left| \rm{{state = active}} \right.} \right)\Pr \left( \rm{{state = active}} \right)\notag\\
= & \frac{1}{2}\left( {1 - v} \right) + \frac{1}{2}\left( {1 - v} \right) + \frac{v}{2}\notag\\
 = & 1 - \frac{v}{2},\tag{B.1}
\end{align}
\begin{align}\label{eqB.2}
&\Pr \left( {\eta  = 1} \right)\notag\\
& = \Pr \left( {A_m(k) = 1\left| \rm{{state = active}} \right.} \right)\Pr \left( \rm{{state = active}} \right)\notag\\
&= \frac{v}{2}.\tag{B.2}
\end{align}

Since $\eta$ takes values in $\left\{ {0,1} \right\}$ and satisfies the probability constraint $\Pr \left( {\eta  = 0} \right)+\Pr \left( {\eta  = 1} \right)=1$, $\eta$ follows a Bernoulli distribution with success parameter $\frac{v}{2}$, $0 \le \frac{v}{2} \le \frac{1}{2}$.

\section*{Appendix C}
Based on (\ref{eq26}), it is easy to obtain the mean square of ${T_{m,k}^{\rm{iq}}}$ under different values of $B_m(k)$ and $B_m(k-1)$. Given equiprobable transmitted bits, the probability of ``00", ``01", ``10", ``11" are $\frac{1}{4}$, the mean and variance of ${T_{m,k}^{\rm{iq}}}$ read
\begin{equation}
\mathbb{E}\left( {T_{m,k}^{\rm{iq}}} \right) = 0,
\tag{C.1}
\label{eqC.1}
\end{equation}
\begin{align}\label{eqC.2}
\mathbb{D}\left( {T_{m,k}^{\rm{iq}}} \right) &= \mathbb{E}\left( {{{\left( {T_{m,k}^{\rm{iq}}} \right)}^2}} \right) - {\mathbb{E}^2}\left( {T_{m,k}^{\rm{iq}}} \right)\notag\\
&= \frac{{{{\left( {\vartheta _{m,k}^{{\rm{iq}}}} \right)}^2}}}{2} + \Delta _{ + }^2.\tag{C.2}
\end{align}

Therefore, $\left| T_{m,k}^{\rm{iq}} \right|$ follows the folded normal distribution\cite{leone1961folded}, with its mean given in (\ref{eqC.3}).


\vspace*{-5pt}

\ifCLASSOPTIONcaptionsoff
  \newpage
\fi
\bibliographystyle{IEEEtran}
\bibliography{refa}

\begin{thebibliography}{10}
\providecommand{\url}[1]{#1}
\csname url@samestyle\endcsname
\providecommand{\newblock}{\relax}
\providecommand{\bibinfo}[2]{#2}
\providecommand{\BIBentrySTDinterwordspacing}{\spaceskip=0pt\relax}
\providecommand{\BIBentryALTinterwordstretchfactor}{4}
\providecommand{\BIBentryALTinterwordspacing}{\spaceskip=\fontdimen2\font plus
\BIBentryALTinterwordstretchfactor\fontdimen3\font minus
  \fontdimen4\font\relax}
\providecommand{\BIBforeignlanguage}[2]{{%
\expandafter\ifx\csname l@#1\endcsname\relax
\typeout{** WARNING: IEEEtran.bst: No hyphenation pattern has been}%
\typeout{** loaded for the language `#1'. Using the pattern for}%
\typeout{** the default language instead.}%
\else
\language=\csname l@#1\endcsname
\fi
#2}}
\providecommand{\BIBdecl}{\relax}
\BIBdecl

\bibitem{345}
Z.~Cui, G.~Wang, M.~Liu, B.~Ai, T.~Q.~S. Quek, and C.~Tellambura, ``Wavy
  signals and striped constellations for backscatter communications: Origins
  and solutions,'' \emph{IEEE Trans. Wireless Commun.}, vol.~23, no.~10, pp.
  12\,815--12\,829, 2024.

\bibitem{10130082}
T.~Jiang, Y.~Zhang, W.~Ma, M.~Peng, Y.~Peng, M.~Feng, and G.~Liu, ``Backscatter
  communication meets practical battery-free internet of things: A survey and
  outlook,'' \emph{IEEE Commun. Surv. Tutorials}, vol.~25, no.~3, pp.
  2021--2051, 2023.

\bibitem{8368232}
R.~Xu, Y.~Ye, H.~Sun, L.~Shi, and G.~Lu, ``Revolutionizing symbiotic radio:
  Exploiting trade-offs in hybrid active-passive communications,'' \emph{IEEE
  Commun. Mag.}, vol.~63, no.~9, pp. 156--163, 2025.

\bibitem{11005959}
F.~Wu, S.~Wang, J.~Zhou, H.~Pan, C.~Zhou, W.~Yang, F.~Lyu, and Y.~Zhang,
  ``Multi-variate time series prediction of traffic and users for dynamic
  {RRH-BBU} mapping in {C-RAN},'' \emph{IEEE Trans. Mob. Comput.}, vol.~24,
  no.~10, pp. 10\,557--10\,572, 2025.

\bibitem{10764739}
X.~Wang, K.~Tao, N.~Cheng, Z.~Yin, Z.~Li, Y.~Zhang, and X.~Shen, ``{RadioDiff}:
  An effective generative diffusion model for sampling-free dynamic radio map
  construction,'' \emph{IEEE Trans. Cognit. Commun. Networking}, vol.~11,
  no.~2, pp. 738--750, 2025.

\bibitem{10353962}
S.~Zargari, A.~Hakimi, F.~Rezaei, C.~Tellambura, and A.~Maaref, ``Signal
  detection in ambient backscatter systems: Fundamentals, methods, and
  trends,'' \emph{IEEE Access}, vol.~11, pp. 140\,287--140\,324, 2023.

\bibitem{liu2013ambient}
V.~Liu, A.~Parks, V.~Talla, S.~Gollakota, D.~Wetherall, and J.~R. Smith,
  ``Ambient backscatter: Wireless communication out of thin air,'' in
  \emph{Proc. ACM SIGCOMM}, Aug. 2013, pp. 39--50.

\bibitem{lu2015signal}
K.~Lu, G.~Wang, F.~Qu, and Z.~Zhong, ``Signal detection and {BER} analysis for
  {RF}-powered devices utilizing ambient backscatter,'' in \emph{Proc. IEEE
  Int. Conf. Wireless Commun. Signal Process (WCSP)}, Oct. 2015, pp. 1--5.

\bibitem{8007328}
J.~Qian, F.~Gao, G.~Wang, S.~Jin, and H.~Zhu, ``Semi-coherent detection and
  performance analysis for ambient backscatter system,'' \emph{IEEE Trans.
  Commun.}, vol.~65, no.~12, pp. 5266--5279, 2017.

\bibitem{wang2016ambient}
G.~Wang, F.~Gao, R.~Fan, and C.~Tellambura, ``Ambient backscatter communication
  systems: Detection and performance analysis,'' \emph{IEEE Trans. Commun.},
  vol.~64, no.~11, pp. 4836--4846, 2016.

\bibitem{7769255}
J.~Qian, F.~Gao, G.~Wang, S.~Jin, and H.~Zhu, ``Noncoherent detections for
  ambient backscatter system,'' \emph{IEEE Trans. Wireless Commun.}, vol.~16,
  no.~3, pp. 1412--1422, 2017.

\bibitem{8329444}
Q.~Tao, C.~Zhong, H.~Lin, and Z.~Zhang, ``Symbol detection of ambient
  backscatter systems with manchester coding,'' \emph{IEEE Trans. Wireless
  Commun.}, vol.~17, no.~6, pp. 4028--4038, 2018.

\bibitem{9242274}
S.~Guruacharya, X.~Lu, and E.~Hossain, ``Optimal non-coherent detector for
  ambient backscatter communication system,'' \emph{IEEE Trans. Veh. Technol.},
  vol.~69, no.~12, pp. 16\,197--16\,201, 2020.

\bibitem{9430725}
W.~Liu, S.~Shen, D.~H.~K. Tsang, and R.~Murch, ``Enhancing ambient backscatter
  communication utilizing coherent and non-coherent space-time codes,''
  \emph{IEEE Trans. Wireless Commun.}, vol.~20, no.~10, pp. 6884--6897, 2021.

\bibitem{9463672}
M.~Ouroutzoglou, G.~Vougioukas, G.~N. Karystinos, and A.~Bletsas, ``Multistatic
  noncoherent linear complexity miller sequence detection for {Gen2}
  {RFID/IoT},'' \emph{IEEE Trans. Wireless Commun.}, vol.~20, no.~12, pp.
  8067--8080, 2021.

\bibitem{angerer2010rfid}
C.~Angerer, R.~Langwieser, and M.~Rupp, ``{RFID} reader receivers for physical
  layer collision recovery,'' \emph{IEEE Trans. Commun.}, vol.~58, no.~12, pp.
  3526--3537, 2010.

\bibitem{wang2018stackelberg}
W.~Wang, D.~T. Hoang, D.~Niyato, P.~Wang, and D.~I. Kim, ``Stackelberg game for
  distributed time scheduling in {RF}-powered backscatter cognitive radio
  networks,'' \emph{IEEE Trans. Wireless Commun.}, vol.~17, no.~8, pp.
  5606--5622, 2018.

\bibitem{gu2022many}
B.~Gu, D.~Li, Y.~Xu, C.~Li, and S.~Sun, ``Many a little makes a mickle: Probing
  backscattering energy recycling for backscatter communications,'' \emph{IEEE
  Trans. Veh. Technol.}, vol.~72, no.~1, pp. 1343--1348, 2022.

\bibitem{li2019price}
D.~Li and Y.-C. Liang, ``Price-based bandwidth allocation for backscatter
  communication with bandwidth constraints,'' \emph{IEEE Trans. Wireless
  Commun.}, vol.~18, no.~11, pp. 5170--5180, 2019.

\bibitem{7420754}
A.-A.~A. Boulogeorgos, V.~M. Kapinas, R.~Schober, and G.~K. Karagiannidis,
  ``{I/Q}-imbalance self-interference coordination,'' \emph{IEEE Trans.
  Wireless Commun.}, vol.~15, no.~6, pp. 4157--4170, 2016.

\bibitem{4357467}
L.~Anttila, M.~Valkama, and M.~Renfors, ``Circularity-based {I/Q} imbalance
  compensation in wideband direct-conversion receivers,'' \emph{IEEE Trans.
  Veh. Technol.}, vol.~57, no.~4, pp. 2099--2113, 2008.

\bibitem{1495892}
A.~Tarighat and A.~Sayed, ``{MIMO} {OFDM} receivers for systems with {IQ}
  imbalances,'' \emph{IEEE Trans. Signal Process.}, vol.~53, no.~9, pp.
  3583--3596, 2005.

\bibitem{7463533}
A.-A.~A. Boulogeorgos, N.~D. Chatzidiamantis, and G.~K. Karagiannidis, ``Energy
  detection spectrum sensing under {RF} imperfections,'' \emph{IEEE Trans.
  Commun.}, vol.~64, no.~7, pp. 2754--2766, 2016.

\bibitem{8085119}
J.~Wang, H.~Yu, Y.~Wu, F.~Shu, J.~Wang, R.~Chen, and J.~Li, ``Pilot
  optimization and power allocation for {OFDM}-based full-duplex relay networks
  with {IQ}-imbalances,'' \emph{IEEE Access}, vol.~5, pp. 24\,344--24\,352,
  2017.

\bibitem{9415632}
S.~Abdallah, A.~I. Salameh, and M.~Saad, ``Joint channel, carrier frequency
  offset and {I/Q} imbalance estimation in ambient backscatter communication
  systems,'' \emph{IEEE Commun. Lett.}, vol.~25, no.~7, pp. 2250--2254, 2021.

\bibitem{10005249}
Y.~Ye, J.~Zhao, X.~Chu, S.~Sun, and G.~Lu, ``Symbol detection of ambient
  backscatter communications under {IQ} imbalance,'' \emph{IEEE Trans. Veh.
  Technol.}, vol.~72, no.~5, pp. 6862--6867, 2023.

\bibitem{schenk2008rf}
T.~Schenk, \emph{{RF} imperfections in high-rate wireless systems: impact and
  digital compensation}.\hskip 1em plus 0.5em minus 0.4em\relax Springer
  Science \& Business Media, 2008.

\bibitem{jalali2019cognitive}
F.~Jalali and A.~Zaimbashi, ``Cognitive radio spectrum sensing under joint
  {TX/RX} {I/Q} imbalance and uncalibrated receiver,'' \emph{IEEE Syst. J.},
  vol.~14, no.~1, pp. 105--112, 2019.

\bibitem{9866050}
Y.~Ye, L.~Shi, X.~Chu, G.~Lu, and S.~Sun, ``Mutualistic cooperative ambient
  backscatter communications under hardware impairments,'' \emph{IEEE Trans.
  Commun.}, vol.~70, no.~11, pp. 7656--7668, 2022.

\bibitem{5740629}
A.~A. El-Sherif and K.~J.~R. Liu, ``Joint design of spectrum sensing and
  channel access in cognitive radio networks,'' \emph{IEEE Trans. Wireless
  Commun.}, vol.~10, no.~6, pp. 1743--1753, 2011.

\bibitem{8103807}
G.~Yang, Y.-C. Liang, R.~Zhang, and Y.~Pei, ``Modulation in the air:
  Backscatter communication over ambient {OFDM} carrier,'' \emph{IEEE Trans.
  Commun.}, vol.~66, no.~3, pp. 1219--1233, 2018.

\bibitem{papoulis2002probability}
A.~Papoulis and S.~U. Pillai, \emph{Probability, Random Variables and
  Stochastic Processes}, 4th~ed.\hskip 1em plus 0.5em minus 0.4em\relax New
  York, NY, USA: McGraw-Hill, 2002.

\bibitem{6746245}
A.~Gokceoglu, S.~Dikmese, M.~Valkama, and M.~Renfors, ``Energy detection under
  {IQ} imbalance with single- and multi-channel direct-conversion receiver:
  Analysis and mitigation,'' \emph{IEEE J. Select. Areas Commun.}, vol.~32,
  no.~3, pp. 411--424, 2014.

\bibitem{leone1961folded}
F.~C. Leone, L.~S. Nelson, and R.~Nottingham, ``The folded normal
  distribution,'' \emph{Technometrics}, vol.~3, no.~4, pp. 543--550, 1961.

\end{thebibliography}

\end{document}